\documentclass[journal]{IEEEtran}
\usepackage[stable]{footmisc}   % Optional: For additional footnote handling
\usepackage{textcomp}
\usepackage{color,array,amsthm, soul}
\usepackage{graphicx}
\usepackage{subcaption}
\usepackage{stfloats}
\usepackage[table,xcdraw]{xcolor}

\usepackage{float}
\usepackage{tcolorbox} % Required for the colored box

\usepackage{caption}
\usepackage{cite} % need to add this so that contiguous references are shown with hyphen
\usepackage{bm}
\usepackage{algorithmic}
\usepackage{color}  % To include comments in color
\usepackage{booktabs}

\usepackage{footnote} % Required for footnotes in tables
\makesavenoteenv{table*} % Allows \footnote inside table* environment

\newcommand{\cp}[1]{\ifmmode {\mathcal{#1}}\else ${\mathcal{#1}}$\fi}
\newcolumntype{P}[1]{>{\centering\arraybackslash}p{#1}}

\usepackage{nccmath}
\def\cred{\textcolor{black}}

\definecolor{darkgreen}{rgb}{0., 0.4, 0.}
\definecolor{amber}{rgb}{1.0, 0.49, 0.0}
\definecolor{orange}{rgb}{1.0, 0.4, 0.0}

\usepackage{amssymb,amsfonts}

\usepackage{algorithm2e}
\usepackage{mathtools}
\usepackage{verbatim}

\usepackage{enumitem}
\renewcommand{\thesection}{\Roman{section}}
\renewcommand{\thesubsection}{\thesection.\Alph{subsection}}

\usepackage{multirow}
\usepackage{array} 
\usepackage[toc,acronyms,nopostdot,nogroupskip, nonumberlist]{glossaries-extra}

\MFUhyphentrue % very important! used to capitalize also word after hyphen in the list of acronyms
\glssetcategoryattribute{general}{glossdesc}{firstuc}
\glssetcategoryattribute{abbreviation}{glossdesc}{title}
\glssetcategoryattribute{abbreviation}{glossdesc}{firstuc=false}

\makeatletter
\newglossarystyle{long-initcapsdesc}{%
  \setglossarystyle{long}%
    \renewcommand{\glsnamefont}[1]{\MakeUppercase{##1}}%
    \renewcommand{\glossentry}[2]{%
    \glsentryitem{##1}\glstarget{##1}{\glossentryname{##1}} &
    \protected@edef\thisdesc{\glsentrydesc{##1}}%
    \xcapitalisewords{\thisdesc}\glspostdescription\space ##2\tabularnewline
  }%
}
\makeatother

\makeglossaries

\usepackage{ragged2e}
\usepackage[colorlinks=true, linkcolor=blue, citecolor=blue, urlcolor=blue]{hyperref}
\usepackage{cleveref}
\begin{document}
%
%
%
%https://sampo.website/blog/en/2023/latex-glossaries/
%https://tex.stackexchange.com/questions/197222/formatting-for-glossaries
%
\newabbreviation{BSS}{BSS}{blind source separation}
\newabbreviation{RL}{RL}{reinforcement learning}
\newabbreviation{RJDM}{RJDM}{radar jamming decision-making}
%
%\newabbreviation{LORO}{LORO}{lobe-on-receive-only}
%
\newabbreviation{RPIP}{RPIP}{random pulse initial phases}
\newabbreviation{CP}{CP}{correlated processing}
\newabbreviation{CPI}{CPI}{coherent processing interval}
%
%\newabbreviation{IC}{IC}{interference cancellation}
%
\newabbreviation{MUSIC}{MUSIC}{multiple signal classification}
%
%\newabbreviation{SAM}{SAM}{swept amplitude-modulation}
%
\newabbreviation{AM}{AM}{amplitude-modulation}
\newabbreviation{NP}{NP}{Neyman-Pearson}
%
%\newabbreviation{DBS}{DBS}{Doppler beam sharpening}
%
\newabbreviation{MTT}{MTT}{multiple target tracking}
\newabbreviation{LFM}{LFM}{linear frequency-modulated}
%
%\newabbreviation{LOS}{LOS}{line-of-sight}
%
\newabbreviation[longplural={directions of arrival}]{DOA}{DOA}{direction \lowercase{of} arrival}
\newabbreviation{SPJ}{SPJ}{self-protection jamming}
\newabbreviation{CFAR}{CFAR}{constant false alarm rate}
\newabbreviation{STFT}{STFT}{short-time Fourier transform}
\newabbreviation{CWD}{CWD}{Choi--Williams distribution}
\newabbreviation{SOJ}{SOJ}{stand-off jamming}
\newabbreviation{PPI}{{PPI}}{plan position indicator}
\newabbreviation{GCN}{{GCN}}{graph convolutional network}
%
%\newabbreviation{MLE}{MLE}{maximum likelihood estimator}
%
%\newabbreviation[longplural={neural networks}]{NN}{NN}{neural network}
%
\newabbreviation{EJ}{EJ}{escort jamming}
\newabbreviation{JRC}{JRC}{joint radar-communication}
\newabbreviation{WLS}{WLS}{weighted least squares}
\newabbreviation{ERP}{ERP}{effective radiated power}
%
%\newabbreviation{LTI}{LTI}{linear time-invariant}
%
\newabbreviation[shortplural={KFs}, longplural={Kalman filters}]{KF}{KF}{Kalman filter}
\newabbreviation[shortplural={EKFs}, longplural={extended Kalman filters}]{EKF}{EKF}{extended Kalman filter}
\newabbreviation[shortplural={PFs}, longplural={particle filters}]{PF}{PF}{particle filter}
\newabbreviation[shortplural={UKFs}, longplural={unscented Kalman filters}]{UKF}{UKF}{unscented Kalman filter}
\newabbreviation[shortplural={GSFs}, longplural={Gaussian sum filters}]{GSF}{GSF}{Gaussian sum filter}
\newabbreviation[shortplural={DOFs}, longplural={degrees of freedom}]{DOF}{DOF}{degree of freedom}
\newabbreviation[shortplural={MMFTs}, longplural={micro-motion false targets}]{MMFT}{MMFT}{micro-motion false target}
\newabbreviation[shortplural={TFTs}, longplural={translational false targets}]{TFT}{TFT}{translational false target}
\newabbreviation[shortplural={PRIs}, longplural={pulse repetition interval}]{PRI}{PRI}{pulse repetition interval}
\newabbreviation[shortplural={PTs}, longplural={physical targets}]{PT}{PT}{physical target}
\newabbreviation{FFT}{FFT}{fast Fourier transform}
%
%\newabbreviation{DA}{DA}{data association}
%
\newabbreviation{SSM}{SSM}{state-space model}
\newabbreviation{JSR}{JSR}{jamming-to-signal ratio}
\newabbreviation{JNR}{JNR}{jamming-to-noise ratio}
\newabbreviation{RCS}{RCS}{radar cross-section}
\newabbreviation{CS}{CS}{compressed sensing}
\newabbreviation{AI}{AI}{artificial intelligence}
\newabbreviation{CRB}{CRB}{Cramér-Rao bound}
\newabbreviation{EW}{EW}{electronic warfare}
\newabbreviation{CW}{CW}{continuous-wave}
\newabbreviation{IF}{IF}{intermediate frequency}
\newabbreviation{HS}{HS}{homogeneity and separation scores}
\newabbreviation[shortplural={TOIs}, longplural={targets of interest}]{TOI}{TOI}{target \MakeLowercase{o}f interest}
\newabbreviation[shortplural={ECMs}, longplural={electronic countermeasures}]{ECM}{ECM}{electronic countermeasure}
\newabbreviation[shortplural={ECCMs}, longplural={electronic counter-countermeasures}]{ECCM}{ECCM}{electronic counter-countermeasure}
\newabbreviation[shortplural={FTs}, longplural={false targets}]{FT}{FT}{false target}
\newabbreviation[shortplural={PDs}, longplural={pulse dopplers}]{PD}{PD}{pulse doppler}
%
%\newabbreviation[shortplural={FTGs}, longplural={false target generator}]{FTG}{FTG}{false target generator}
%
\newabbreviation[shortplural={FDAs}, longplural={frequency diverse arrays}]{FDA}{FDA}{frequency diverse array}
\newabbreviation[shortplural={SARs}, longplural={synthetic aperture radars}]{SAR}{SAR}{synthetic aperture radar}
\newabbreviation[shortplural={DRFMs}, longplural={digital radio-frequency memories}]{DRFM}{DRFM}{digital radio frequency memory}
\newabbreviation[shortplural={RGPOs}, plural={range gate pull-offs}]{RGPO}{RGPO}{range gate pull-off}
\newabbreviation[shortplural={RGPIs}, longplural={range gate pull-ins}]{RGPI}{RGPI}{range gate pull-in}
\newabbreviation[shortplural={VGPOs}, longplural={velocity gate pull-offs}]{VGPO}{VGPO}{velocity gate pull-off}
\newabbreviation[shortplural={VGPI}s, longplural={velocity gate pull-ins}]{VGPI}{VGPI}{velocity gate pull-in}
%
%\newabbreviation[shortplural={RVGPOs}, longplural={range-velocity gate pull-offs}]{RVGPO}{RVGPO}{range-velocity gate pull-off}
%
\newabbreviation[shortplural={RVGPIs}, longplural={range-velocity gate pull-ins}]{RVGPI}{RVGPI}{range-velocity gate pull-in}
\newabbreviation[shortplural={ISRJs}, longplural={interrupted-sampling repeater jammings}]{ISRJ}{ISRJ}{interrupted-sampling repeater jamming}
\newabbreviation{CRDJ}{CRDJ}{crosspulse repeater deception jamming}
\newabbreviation[shortplural={MHTs}, longplural={multiple hypothesis trackings}]{MHT}{MHT}{multiple hypothesis tracking}
\newabbreviation[shortplural={RNNs}, longplural={recurrent neural networks}]{RNN}{RNN}{recurrent neural network}
\newabbreviation[shortplural={SJNRs}, longplural={signal-to-jammer noise ratios}]{SJNR}{SJNR}{signal-to-jammer noise ratio}
\newabbreviation[shortplural={CNNs}, longplural={convolutional neural networks}]{CNN}{CNN}{convolutional neural network}
\newabbreviation[shortplural={LSTMs}, longplural={long short-term memories}]{LSTM}{LSTM}{long short-term memory}
\newabbreviation[shortplural={TDOAs}, longplural={time differences of arrival}]{TDOA}{TDOA}{time difference of arrival}
\newabbreviation[shortplural={RFSs}, longplural={random finite sets}]{RFS}{RFS}{random finite set}
\newabbreviation[shortplural={SNRs}, longplural={signal-to-noise ratios}]{SNR}{SNR}{signal-to-noise ratio}
\newabbreviation[shortplural={SJRs}, longplural={signal-to-jammer ratios}]{SJR}{SJR}{signal-to-jammer ratio}
\newabbreviation[shortplural={OFDMs}, longplural={orthogonal frequency-division multiplexings}]{OFDM}{OFDM}{orthogonal frequency-division multiplexing}
\newabbreviation[shortplural={PRFs}, longplural={pulse repetition frequencies}]{PRF}{PRF}{pulse repetition frequency}
\newabbreviation[shortplural={SIMOs}, longplural={single-input multiple-outputs}]{SIMO}{SIMO}{single-input multiple-output}
\newabbreviation[shortplural={MIMOs}, longplural={multiple-input multiple-outputs}]{MIMO}{MIMO}{multiple-input multiple-output}
\newabbreviation[shortplural={GLRTs}, longplural={generalized likelihood ratio tests}]{GLRT}{GLRT}{generalized likelihood ratio test}
\newabbreviation[shortplural={AGCs}, longplural={automatic gain controls}]{AGC}{AGC}{automatic gain control}
\newabbreviation[shortplural={MLs}, longplural={machine learnings}]{ML}{ML}{machine learning}
\newabbreviation[shortplural={UAVs}, longplural={unmanned aerial vehicles}]{UAV}{UAV}{unmanned aerial vehicle}
\newabbreviation[shortplural={GNSSs}, longplural={global navigation satellite systems}]{GNSS}{GNSS}{global navigation satellite systems}
% Include the acronyms file
\title{Advances in Anti-Deception Jamming Strategies for \\Radar Systems: A Survey}
%in Electronic Warfare:
% Other title options:
% - Anti-Deception Jamming Strategies in Radar Systems: Trends, Challenges, and Advances
% - Advances in Anti-Deception Jamming Strategies in Radar Systems: A Survey
% - Anti-Deception Jamming Strategies for Radar Systems: A Comprehensive Review
%
\author{Helena Calatrava,~\IEEEmembership{Student Member,~IEEE}, Shuo Tang,~\IEEEmembership{Student Member,~IEEE}, \\ and Pau Closas,~\IEEEmembership{Senior Member,~IEEE}
%\affil{Northeastern University, Boston, MA 02115, USA} 
%

%\author{Pau Closas}
%\member{Senior Member, IEEE}
%\affil{Northeastern University, Boston, MA 02115, USA} 
%
%% \author{FOURTH D. AUTHOR}
%% \affil{University of Colorado, Colorado, USA}
%
\thanks{Manuscript received XXXXX 00, 0000; revised XXXXX 00, 0000; accepted XXXXX 00, 0000.}
\thanks{This work has been partially supported by the National Science Foundation under Awards ECCS-1845833 and CCF-2326559. {\itshape (Corresponding author: Helena Calatrava)}.}
%
%\thanks{{\itshape (Corresponding author: Helena Calatrava)}.}
%
%
\thanks{Helena Calatrava, Shuo Tang and Pau Closas are with Northeastern University, Boston, MA 02115, USA
(e-mail: \{calatrava.h,\ tang.shu,\ closas\}@northeastern.edu).}}
%
%\supplementary{Code and demonstration of experimental results in video format are available at XXXXX}
%\supplementary{Code and demonstration of experimental results in video format are available at \href{http://ieeexplore.ieee.org}{http://ieeexplore.ieee.org}.}
%
%
\markboth{CALATRAVA ET AL.}{ADVANCES IN ANTI-DECEPTION JAMMING STRATEGIES FOR RADAR SYSTEMS}
\maketitle

%
%
%\vspace{-0.39cm}
\begin{abstract}
Deception jamming has long been a significant threat to radar systems, interfering with search, acquisition, and tracking by introducing false information that diverts attention from the targets of interest.
As deception strategies become more sophisticated, the vulnerability of radar systems to these attacks continues to escalate.
This paper offers a comprehensive review of the evolution of anti-deception jamming techniques, starting with legacy solutions and progressing to the latest advancements.
Current research is categorized into three key areas: prevention strategies, which hinder the ability of jammers to alter radar processing; detection strategies, which alert the system to deception and may classify the type of attack; and mitigation strategies, which aim to reduce or suppress the impact of jamming.
%
%Mitigation techniques are further divided into signal domain mitigation, which suppresses deceptive measurements at the signal level, and measurement domain mitigation, which addresses the effects of deceptive measurements on state estimation.
%
Additionally, key avenues for further research are highlighted, with a particular emphasis on distributed, cognitive, and AI-enabled radar systems.
%
%This includes the integration of Transformer-based models and reinforcement learning.
%
We envision this paper as a gateway to the existing literature on anti-deception jamming, a critical area for safeguarding radar systems against evolving threats. 
%Our hope is that it will equip both practitioners and researchers with the necessary insights to implement current solutions and contribute to advancing the field.
%
%Finally, we explore key avenues for further research, with a particular focus on distributed, cognitive, and AI-enabled radar systems, which includes hte integration of Transformer-based models and reinforcement learning.
%
%Our hope with this paper is to provide both practitioners and theorists with valuable insights to not only deploy existing solutions but also drive innovation in this critical field, essential for protecting radar systems against increasingly sophisticated threats.
\end{abstract}


\renewcommand{\glsnamefont}[1]{\makefirstuc{#1}}
%\printglossary[title={List of Abbreviations}]
%\printglossary[style=super, title={foo}[title={foo}]]
%
% SECTION: INTRODUCTION
\section{Introduction}\label{sec:intro}
%
%\helen{daniel chu comments about drone interdiction (check drone study center)}
%
%Deception jammers, also referred to as repeater jammers, are employed in radar-dense environments to alter the ability of an adversary to detect, identify, and track \glspl{PT} in the scene. 
%
%In self-protection scenarios, the jammer assumes the role of the \gls{TOI} and uses deception techniques to remain undetected by the adversarial radar.
%
%\Gls{ECM} systems represent a specialized subset of \gls{EW} designed to degrade the effective use of the electromagnetic spectrum by an adversary~\cite{butt2013overview}.
%
%One common use case of \gls{EW} involves disrupting radar lock-on to protect aircraft from missile-guided threats.
%
%As a fundamental component of \gls{ECM}, deception jammers replicate the radar waveform to mislead radar search, acquisition, or tracking~\cite{cheng2022introduction}, either by introducing false information or by generating multiple \glspl{FT} in rapid succession to induce confusion or overload the system.
%
%Two illustrative examples of deception jamming strategies are shown in Fig.~\ref{fig:intro:deception_jamming_scenario}.
%

\begin{figure}
\centering
  \includegraphics[width=1\columnwidth]{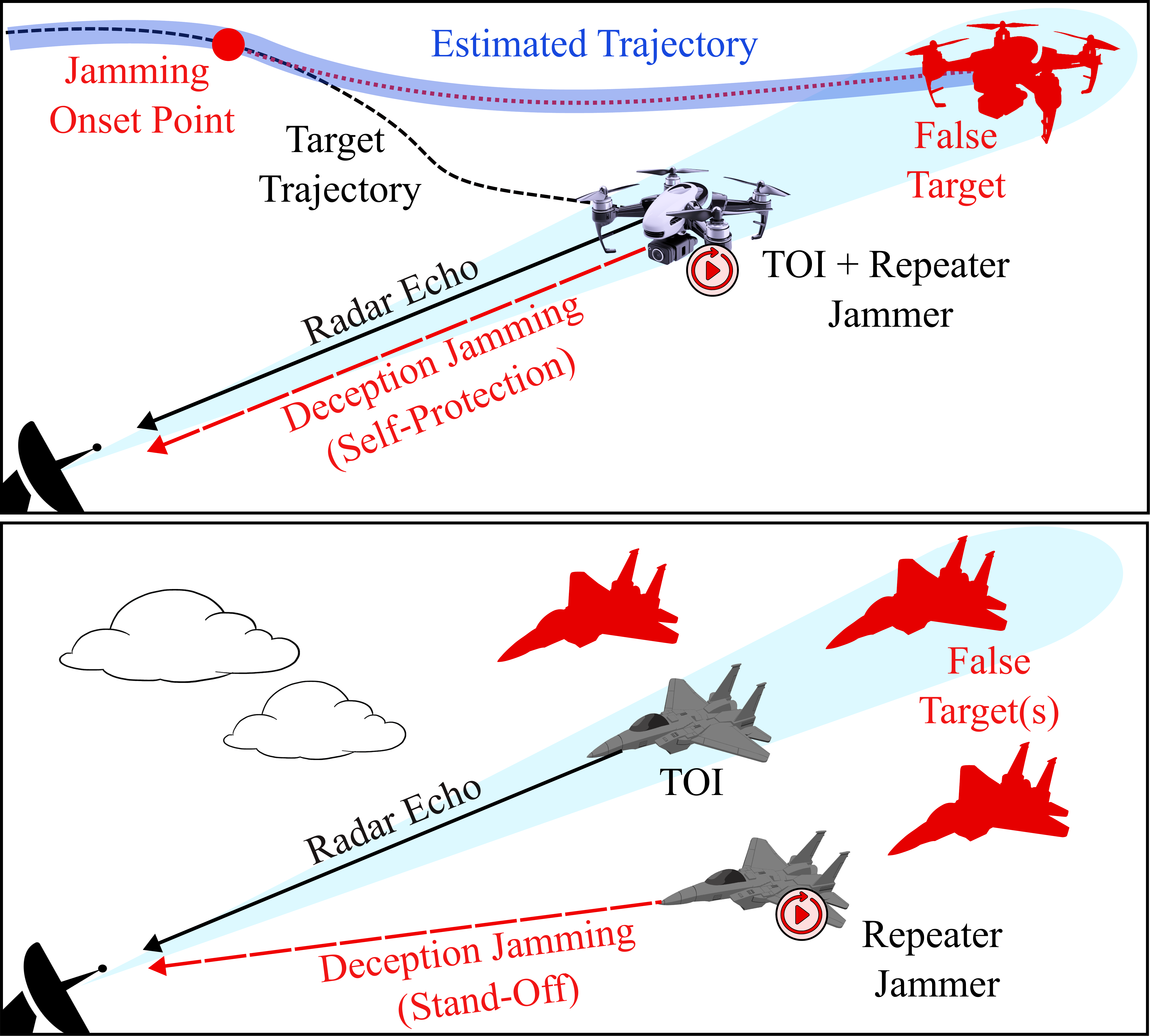}
  \caption{Illustration of deception jamming effects in radar signal reception: (Top) Tracking deception misleads the radar into estimating a false target trajectory. (Bottom) Multiple false targets are generated to hinder target detection.}
  \label{fig:intro:deception_jamming_scenario}
  %\vspace{-.8cm}
\end{figure}

\Gls{ECM} systems represent a subset of \gls{EW} designed to degrade the effective use of the electromagnetic spectrum by an opponent~\cite{graham2011communications, butt2013overview}.
As a fundamental component of \gls{ECM}, deception jammers, also known as repeater jammers, are employed in radar-dense environments to alter the ability of an adversarial radar to detect, identify, and track \glspl{PT} in the scene~\cite{neri2006introduction}.  
By replicating the radar waveform, these jammers mislead radar search, acquisition, or tracking, either by introducing false information or rapidly generating multiple \glspl{FT} to overload the processing capabilities of the radar~\cite{kwak2009application}.
One common use case of deception jamming involves disrupting radar lock-on to protect aircraft from missile-guided threats~\cite{cheng2022introduction}.
In self-protection scenarios, the jammer assumes the role of the \gls{TOI} and uses deception strategies to remain undetected~\cite{Hanbali_Kastantin_2017}.
Two illustrative examples of deception jamming strategies are shown in Fig.~\ref{fig:intro:deception_jamming_scenario}.

Notably, the jamming signal can generate an \gls{FT} without exceeding the power of the \gls{TOI} echo, as long as it surpasses the radar detection threshold~\cite{ew_101}.
Unlike noise jammers, deception jammers do not transmit during the entire duty cycle of the radar signal, which improves power efficiency and reduces system weight. Additionally, their greater concealment decreases the likelihood of detection by adversarial \cred{systems.
Despite} these advantages, deception jammers require high memory capacity and sensitivity to accurately track and replicate radar echoes~\cite{neri2006introduction}.
A comprehensive understanding of the characteristics and processing strategies of deception attacks is essential for developing advanced \gls{ECCM} strategies that enhance radar resilience in the presence of such threats~\cite{adamy2015ew}.
In this context, this survey explores advances in anti-deception jamming to counter increasingly sophisticated electronic attacks.
%
%This paper explores anti-deception jamming strategies aimed at ensuring radar systems can accurately track PTs amidst electronic interference.

%In \cite{neri2006introduction}, various radar sensors, including search/surveillance radar, Synthetic Aperture Radar (SAR), tracking radars, and airborne radars, are discussed within the context of electronic warfare (EW).

\begin{figure*}
\centering
  \includegraphics[width=.95\textwidth]{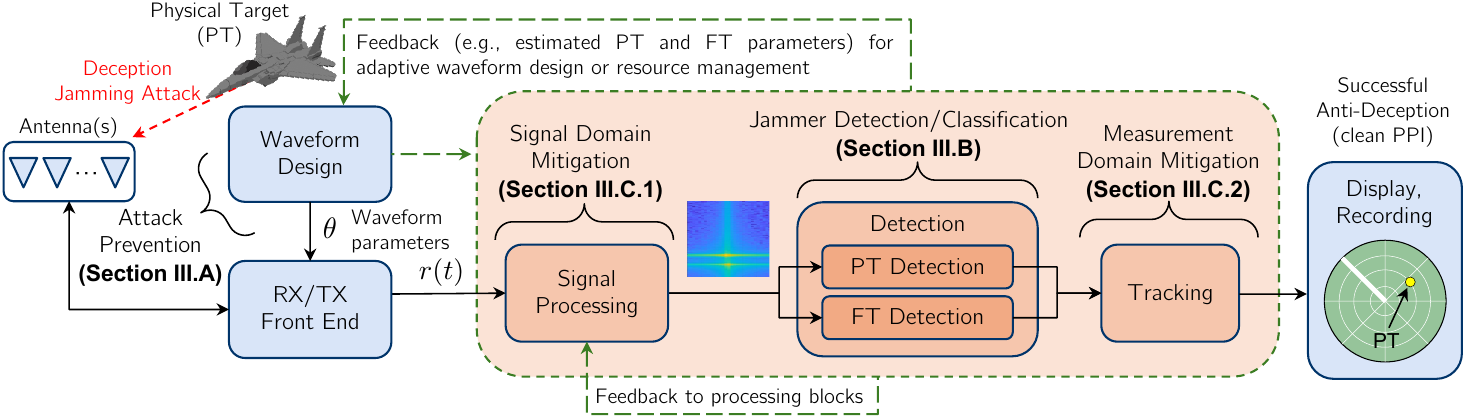}
  %\vspace{.1cm}
  \caption{\cred{Overview of the radar signal processing chain highlighting principal functional blocks, each mapped to the pertinent section of this survey. The displayed \gls{PPI} represents a clean (protected) radar output resulting from the successful application of anti-deception jamming strategies, facilitating reliable identification of the \gls{PT}. For a detailed classification of methods within each radar processing stage, see Fig.~\ref{fig:taxonomy_jamming_iii}.}}
   \label{fig:intro}
  % \vspace{-2.5ex}
\end{figure*}

\setlength{\arrayrulewidth}{0.1mm} % optional: adjust the thickness of table borders
\renewcommand{\arraystretch}{1.5}  % optional: adjust row height
\begin{table*}[h]
    \centering
    \caption{\cred{Overview of the tables in this survey presenting a review of radar anti-deception jamming works.}}
    \begin{tabular}{|c|c|P{13cm}|} % Now centers the Description column
        \hline
        \rowcolor{blue!10}
        \textbf{Table} & \textbf{Anti-Deception Function} & \textbf{Description} \\ \hline
        \ref{table:jamming_prevention} & Prevention & Strategies using pulse diversity to counter deception jamming by modifying radar signal parameters. \\ \hline
        \ref{table:jamming_discrimination} & Detection & Decision-making strategies for detecting, discriminating, and/or classifying deception jamming. \\ \hline
        \ref{table:detection_multistatic} & Detection & Multistatic radar techniques for deception jamming detection, presented separately from Table~\ref{table:jamming_discrimination} due to their significant presence in the literature. \\ \hline
        \ref{table:spatial-frequency diversity} & Mitigation & Mitigation strategies using spatial (multistatic radar) and spatial-frequency (FDA-MIMO radar) diversity. \\ \hline
    \end{tabular}
    \label{table:summary_of_tables}
\end{table*}

%We begin by introducing the concept of coherence, as it plays a crucial role in radar performance and countermeasure effectiveness.
%
We begin by introducing coherent radar systems, which depend on precise synchronization of signal phase and frequency to enhance detection and processing capabilities, whereas non-coherent radar systems operate without this requirement~\cite{skolnik1970radar}. The development of coherent signal processing techniques in the 1970s, including pulse compression, \gls{PD} radar, and \gls{SAR}, posed a challenge to the traditional repeater jammers prevailing at the time. Although these jammers were capable of techniques such as gate stealing and spoofing, they were unable to exploit the processing gains associated with signal coherence~\cite{neri2006introduction}.
%due to their lack of coherence with the true target echo.
%
This limitation led to the development of \gls{DRFM} technology in the 1990s~\cite{roome1990digital}, enabling precise monitoring, storage, modification of signal parameters such as delays or Doppler shifts, and nearly perfect replay of radar signals~\cite{schleher1986introduction}.
Although conceptually simple, \gls{DRFM} devices are technologically complex due to the high-speed digital processing they require. Under direct computer control, they are capable of both coherent and incoherent jamming. The concept of coherence in jamming is introduced in Section~\ref{sec:jamming_general_taxonomy}.
%
%Unlike coherent techniques, non-coherent jamming does not benefit from the processing gains associated with signal coherence.
%, including the deception attacks described in~\Cref{sec:jamming_deception_taxonomy}, which are the focus of this survey.
%Traditionally, radar systems assumed that electronic countermeasures could not track their operating frequency or match their sampling rate. --> makes no sense to mention this again because this is related to synchronization
%
%

Advances in signal processing technologies, such as high-speed sampling and the replication of wideband radar signals~\cite{richards2005fundamentals}, underscore the critical need for evolving anti-deception strategies to counter increasingly sophisticated \gls{DRFM} jamming threats~\cite{sparrow2006ecm, kwak2009application}.
Essentially, \gls{ECM} and \gls{ECCM} evolve in tandem, with advances in each driving progress in the other, both fueled by the rapid growth of computer hardware~\cite{spezio2002electronic}.
Nevertheless, while this leads to implementation methods that are continually evolving, the principles underlying deception attacks generally remain central~\cite{ew_101}. An example of this can be seen with \gls{RGPO} attacks~\cite{wang2024modelling}, which were initially non-coherent, later incorporating coherence, and today using optimization techniques to maximize their deception success rate~\cite{jia_intelligent_2020}.
%
%While the evolution of deception attacks is clearly driven by technological advances, their core concept remains central, e.g., range gate stealing in the case of \glspl{RGPO}. 
%
This underscores the importance of revisiting classical techniques to understand the motivations and foundational principles that shape modern countermeasures and their challenges.
To this end, in this paper we review the literature on countermeasures against radar deception attacks, starting in the first decade of the 2000s~\cite{1435879, akhtar_eccm_2007, 4203039, 4472184, schuerger2008deception, akhtar2009orthogonal, schuerger2009performance} to the present, and culminating in a discussion \cred{on the role of \gls{ML} and broader \gls{AI} approaches}~\cite{kang2018reinforcement, sharma2020artificial, luo2021semi, lv2021radar, kong2022active, cheng2022introduction, bouzabia2022deep, jiang2023intelligent, wang2023application, peng2023positive, wenbin2024method, zhu2024gcn,yang2024hybrid,fi15120374,10868277, wrabel2021a}, cognition (including game theory)~\cite{ wen2019cognitive, 7485306, ge2021joint,zhang2023cognitive, wang2024design, zhang2019game, nan2020mitigation, he2021game, geng2023radar, ibrahim2024game, he2024game, zhang2024game}, and networked along with distributed radar architectures~\cite{zhao2015signal, zhao2017discrimination, yu2019polarimetric, yu2021message, zhang2022joint, zhao2022cooperative, chalise2022distributed, yang_consensus_2023, sun2024anti, zhang2024game, sun2024coordinated, rs16142616}.

\textbf{Related Surveys:}
The literature on radar anti-deception jamming encompasses a wide range of approaches with varying assumptions and objectives, making systematic comparison challenging.
To better understand the current landscape, we review existing surveys before highlighting what distinguishes our study.
The survey in~\cite{395232} provides a strong theoretical foundation on \gls{ECM}/\gls{ECCM} but does not address recent advances in the field.
The overview in~\cite{butt2013overview} introduces \gls{EW} strategies such as frequency hopping and pulse compression but does not extensively explore the literature, while~\cite{Hanbali_Kastantin_2017} specifically focuses on deception attacks against chirp radars.
Recent surveys provide valuable insight into state-of-the-art methods but tend to focus on specific research directions.
Among these, the work in~\cite{electronics11193025} examines deep learning applications for \glspl{UAV}, addressing challenges such as cyberattacks~\cite{10261240} and \gls{GNSS} spoofing, which fall outside the scope of our study as we focus on radar systems.
Additionally, \gls{AI}-based \gls{EW} strategies and cognitive jamming decision-making approaches are reviewed in~\cite{sharma2020artificial} and \cite{10106133}, respectively. Another recent survey covers game-theoretic anti-jamming techniques but is specific to cognitive radio networks and does not address the \gls{FT}-generation capabilities of deception jammers~\cite{ibrahim2024game}.
A broader survey on the role of game theory in defense systems is presented in~\cite{ho2022game}.
Finally, emerging trends in metacognitive radar are reviewed in~\cite{martone_2025_metacognitive}.

%
%
%

%

%

%
%

%We also bring out the interrelationship between different methods, wherever possible. Issues related to the implementation of methods are out of this paper.

%\hl{Highlight the limitations or gaps in current classifications or taxonomies. Mention previous surveys/similar surveys and explain why ours is going to be relevant to the radar community.}
%

%\textcolor{orange}{Prevention and detection strategies operate primarily in the signal domain, whereas mitigation strategies can extend into the measurement domain, where the focus shifts to reducing the impact of deceived measurements, such as range, Doppler, and \gls{DoA}.}

\begin{figure*}[t]
%\hspace{-.5cm}
% \includegraphics[width=0.8\textwidth, height=3.8cm]{figures_TAES_FREEDOM/results/chaotic/chaotic_rmse.pdf}
\centering
  \includegraphics[width=1\textwidth]{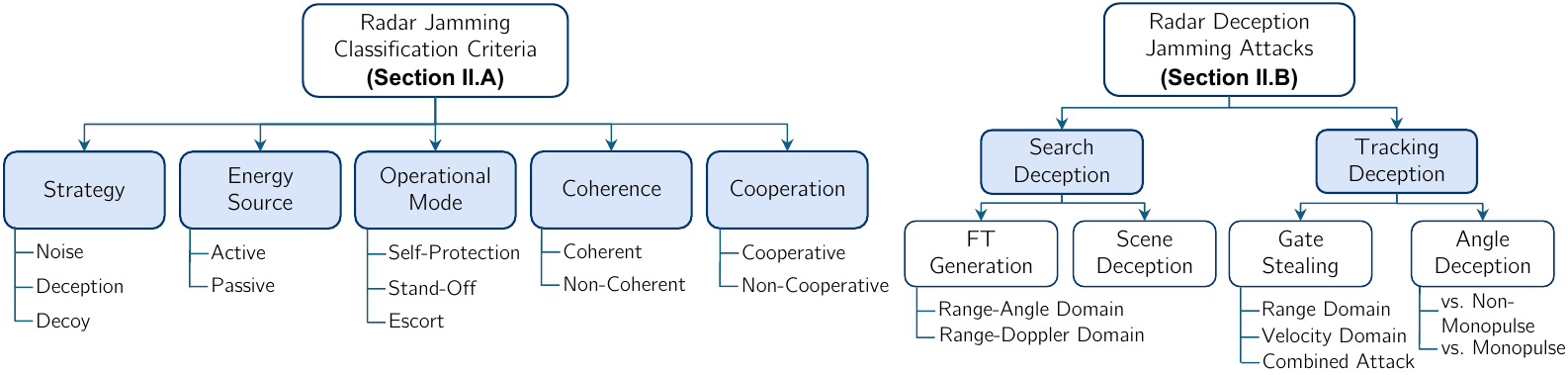}
  %\vspace{-.05cm}
  \caption{(Left): Classification of jamming attacks based on different criteria, as introduced in Section~\ref{sec:jamming_general_taxonomy}. (Right): Classification of radar deception jamming strategies based on the targeted radar processing stage, as described in~\Cref{sec:jamming_deception_taxonomy}.}
   \label{fig:taxonomy_jammming_ii}
  % \vspace{-2.5ex}
\end{figure*}

%
To the best of the authors’ knowledge, this is the first comprehensive survey that covers both legacy methods and recent advancements in radar anti-deception jamming techniques. These are classified into three main categories based on their operational focus: prevention, detection, and mitigation. A detailed taxonomy reflecting the operational classification is provided in Fig.~\ref{fig:taxonomy_jamming_iii}, \cred{while Fig.~\ref{fig:intro} presents a complementary overview linking these strategies to the corresponding stages of the radar signal processing chain where they are applied.}
%
%In this figure, the role of multistatic radar in the detection and mitigation categories is highlighted. The use of multiple sensors provides spatial diversity, which is crucial in distinguishing between \glspl{FT} and \glspl{PT}, as discussed in Sections~\ref{sec:detection} and~\ref{sec: mitigation}.
%
Table~\ref{table:summary_of_tables} outlines the contents of Tables \ref{table:jamming_prevention}–\ref{table:spatial-frequency diversity}, which cover the majority of the reviewed works, although other pertinent studies are discussed in the text.
Additionally, we provide an overview of jamming attacks, with a specific focus on radar deception jamming, for which the proposed taxonomy is outlined in Fig.~\ref{fig:taxonomy_jammming_ii}.
Building on the limitations and gaps identified in previous surveys, the main contributions of this work are as follows:
%\vspace{-0.1cm}
%
\begin{itemize}
    %\item We present a taxonomy of jamming systems and describe the main types of deception jamming relevant to the anti-jamming strategies discussed later.
    \item We provide a taxonomy for anti-deception jamming in radar systems based on their functional role, discussing solutions for the prevention, detection, and mitigation of deception attacks.
    \item We investigate solutions within each category, tracing their evolution from legacy methods to the development of state-of-the-art technologies.
    \item We summarize the open challenges and future research directions in radar deception jamming countermeasures, with a particular emphasis on emerging technologies such as distributed, cognitive, and AI-enabled radar systems.
\end{itemize}
%
%\hl{Describe our taxonomies and provide diagrams/tables for clarity.}
%\cred{Probably this part of the introduction could substitute Section III, but I am leaving it there because of comfort in writing the first draft.}
%
%

The rest of the paper is organized as follows. \Cref{sec:jamming_in_radar} provides a general introduction to jamming attacks with a particular focus on radar deception. \Cref{sec:taxonomy_anti} provides a comprehensive review and taxonomy of anti-deception jamming strategies for radar systems. \Cref{sec:challenge_future} discusses the main challenges and identifies future research directions in this field. Finally, \Cref{sec:conclusion} concludes the paper with final remarks.

\section{Background on Radar Deception Jamming}\label{sec:jamming_in_radar}
%This section introduces deception jamming strategies against radar systems. 
We begin with an overview of jamming attacks in~\Cref{sec:jamming_general_taxonomy}, followed by a focused discussion on radar deception jamming in~\Cref{sec:jamming_deception_taxonomy}.
For further details on \gls{ECM}, interested readers are directed to~\cite[Ch. 5]{neri2006introduction} and \cite[Ch. 9]{ew_101}.

\subsection{Introduction to Jamming Attacks} \label{sec:jamming_general_taxonomy}

%\begin{figure*}
%\centering
 % \includegraphics[width=0.95\textwidth]{figures/class notes 4.jpeg}
  %\vspace{-.05cm}
  %\caption{\cred{Do we want a figure with deception jamming taxonomy?}}
   %\label{fig:intro:jamming_taxonomy}
%\end{figure*}

%In the literature, jamming is referred to as electronic attack (EA) or electromagnetic countermeasure (ECM)
%, which aims to prevent the correct operation of an enemy's weapon system without resorting to conventional weapons~\cite{neri2006introduction}. 
%
Jamming attacks aim to disrupt the reception capabilities of an adversarial system without causing physical damage, making them a form of ``soft kill'' method. 
%A fundamental principle of jamming is to attack the receiver, not the transmitter, to prevent it from processing intended signals effectively.
%
%
%\textbf{Communications, Radar, and GNSS Jamming:}
%
%Jamming can be classified by the type of signal it targets~\cite[Ch. 9]{ew_101}. Communications jamming disrupts the flow of information between communicating entities, while 
%In particular, radar jamming interferes with the radar's ability to recover information about a target from its return signal. 
%
Jamming can be classified based on the type of signal it targets, such as communications or radar.
As \cred{joint radar-communication} systems become increasingly prevalent, the distinction between communications and radar jamming is diminishing due to shared hardware and frequency bands~\cite{8972666}.
Jammers can also interfere with navigation systems, as in the case of \gls{GNSS} spoofing attacks~\cite{9656691}. These involve transmitting counterfeit satellite signals to induce errors in navigation or timing information~\cite{amin2016vulnerabilities,radovs2024recent}, similar to deception jamming in radar.
%
%Global Navigation Satellite System (GNSS) jamming could also be included in this taxonomy. Similar to communications jamming, it disrupts the reception of signals transmitted by a source (in this case, satellites from a navigation constellation), but this time targeting positioning systems.
%
%As the focus of this paper is on radar attacks, the interested reader is directed to~\cite{radovs2024recent} for a comprehensive survey on GNSS jamming, and to~\cite{pirayesh2022jamming} for a review in the context of wireless networks.
%
%
%We present the following structured overview of jamming attacks.
%Next, we discuss several principles behind radar jamming attacks, including their strategy, energy source, and operational mode, as depicted in Fig.~\ref{fig:intro:jamming_taxonomy_general}.
%
%\cred{Next, we discuss several principles behind radar jamming attacks, including their strategy, energy source, and operational mode, as depicted in Fig.~\ref{fig:intro:jamming_taxonomy_general}.}
Radar jamming attacks can be categorized based on multiple criteria, as depicted in Fig.~\ref{fig:taxonomy_jammming_ii}, and elaborated upon in the following discussion.

\textbf{Noise, Deception, and Decoy Jamming:}
Noise jamming, also referred to as cover jamming, elevates the background noise at the receiver, thereby reducing the \gls{SNR}. When the \gls{SNR} falls below a critical threshold, the ability of the receiver to effectively track or detect the target is compromised.
Types of noise jamming include spot jamming, which affects a limited bandwidth, and wideband noise jamming, such as barrage and sweep jamming, which cover the full bandwidth of a frequency-agile radar.
Deception jamming, on the other hand, intercepts, modifies, and replays the radar signal to introduce misleading information~\cite[Ch. 9]{ew_101}.
We elaborate on the main types of radar deception jamming in~\Cref{sec:jamming_deception_taxonomy}.
In certain scenarios, employing a combination of noise and deception jamming techniques is advantageous~\cite{neri2006introduction}.
An additional jamming strategy, decoy jamming, involves a special type of jammer designed to appear to the opposing radar as more similar to the \gls{TOI} than the \gls{TOI} itself.
%
%\cred{While deception jamming manipulates the radar signal to provide false information, decoy jamming involves creating a false target, either physically or electronically, to divert the radar's attention away from the true target.}
%
%Unlike classical jammers, which interfere with the radar’s operation, a decoy does not directly disrupt radar functionality. 
%
They can be classified into three types~\cite[Ch. 10]{ew_101}: expendable, which are used briefly and discarded; towed, attached by cable to an aircraft or ship for extended protection; and independent maneuver, self-propelled decoys with flexible movement.
The primary functions of these decoys are to overwhelm enemy defenses with the generation of multiple \glspl{FT} (saturation), divert attacks away from the \gls{TOI} (seduction), or provoke the radar into revealing offensive capabilities by responding to a decoy (detection of the adversarial radar).
%
%Although decoys can be used as part of a deception strategy, decoy jamming is not considered a primary deception technique and therefore falls outside the scope of this survey.

\textbf{Active vs. Passive Jamming:}
Active jamming generates electromagnetic energy to disrupt radar operations, which is the case of noise, deception, and some forms of decoy jamming. In contrast, passive jamming relies on methods like the use of confusion reflectors such as chaff~\cite{Yang2018SimulationAO}, or chemical countermeasures including smoke or aerosols~\cite{sharma2020artificial}.
Remarkably, stealth technology, which aims to reduce the visibility of a radar system, can use both passive and active measures. Examples include minimizing the \gls{RCS} with absorbing materials (passive)~\cite{KIM2009161} and the emission of \gls{RCS}-masking signals (active).
Decoys can be passive by replicating a radar signature comparable to the \gls{TOI}, or active by retransmitting a stronger echo or high-power noise.
%Active jamming is more versatile but detectable, while passive jamming is stealthier but less controllable.

\textbf{Onboard vs. Offboard Jamming:}
%
%We define the protected platform as the one seeking to evade detection or tracking, typically operating within the lethal range of enemy weapons. According to their operational mode, 
%
When the jamming source is integrated into the platform that the jammer seeks to protect from radar detection, it is known as an onboard system and is typically associated with \gls{SPJ}.
In contrast, \gls{SOJ} and \gls{EJ} are classified as offboard systems, as they employ a jammer on a separate platform to provide area-wide protection. While \gls{SOJ} operates outside adversarial radar coverage \cred{(as shown in the bottom panel of Fig.~\ref{fig:intro:deception_jamming_scenario})}, \gls{EJ} often operates within it and mirrors the target maneuvers.
%
%Additionally, \gls{EJ} aims to mask the protected platform as it penetrates adversarial territory. The escorting aircraft 
%Also, , \gls{EJ} often operates within the adversarial radar coverage and performs the same maneuvers as the \gls{TOI}. Both .
%
%
%SPJ is also referred to as onboard jamming, while SOJ and EJ are classified as off-board jamming.
%
Deception jamming, while applicable across various operational modes, is generally considered a self-protection mechanism~\cite{neri2006introduction}.
Main-beam deception jamming occurs naturally in \gls{SPJ} mode, where the jammer is co-located with the \gls{TOI} and consequently the \gls{DOA} of the jamming signal aligns with the radar-target \cred{line-of-sight} direction~\cite{wang_main-beam_2020}. \cred{This is illustrated in the top panel of Fig.~\ref{fig:intro:deception_jamming_scenario}.}
%
%Moreover, certain deception techniques are constrained by their geometry. This is the case of \gls{RGPO} attacks, which require the jamming signal to align with the radar-target \gls{LOS}. This alignment occurs naturally in \gls{SPJ} mode, where the jammer is co-located with the \gls{TOI}~\cite{wang_main-beam_2020}.
%and thus, the attack is also referred to as main-beam deception jamming in the literature --> I commented this to reduce text and also because now i am doubting whether this paper mentioned specifically RGPO or main beam deception jamming in general

%\pau{after reading section 2.A, seems like these different jamming types and classifications could be visualy shown in a table or diagram.}

\textbf{Non-Coherent vs. Coherent Jamming:}
Coherent jamming relies on precise synchronization with the radar signal, while incoherent jamming uses unsynchronized noise or broad-spectrum interference. Common forms of incoherent jamming include barrage jamming, spot jamming, and swept jamming. Coherent jamming techniques, such as deception jamming, are more complex but also more efficient, requiring less power to mislead the radar.
%Coherent jamming relies on precise synchronization with the radar signal, while incoherent jamming uses unsynchronized noise or broad-spectrum interference. Common forms of incoherent jamming include barrage jamming, which overwhelms radar with wideband noise across multiple frequencies; spot jamming, which targets a specific frequency or channel; and swept jamming, which extends spot jamming to cover a range of frequencies. Coherent jamming techniques, such as deception jamming, are more complex but also more efficient, requiring less power to mislead radar.
%
There are also simpler coherent methods, such as coherent blink jamming, which switches on and off in sync with the radar \gls{PRI}, creating intermittent interference that causes confusion~\cite{davidson2020theory}.

\textbf{Non-Cooperative vs. Cooperative Jamming:}
%Non-cooperative jamming involves a single jammer acting independently to disrupt radar or communication systems, lacking coordination with other jammers.
%
Cooperative jamming involves a coordinated effort among multiple entities that synchronize their actions and share information to enhance deception.
In the context of distributed radar systems, networked jamming has become increasingly relevant. Examples include UAV swarms for distributed cooperative jamming~\cite{electronics11020184} and cooperative deception jamming power scheduling designed to counter distributed radar networks~\cite{sun2024coordinated}.
Furthermore, the authors in~\cite{zhou2019deception} and~\cite{tian2016multi} propose multi-receiver deception jamming techniques against near-field \gls{SAR}. Both methods use networked receivers to intercept SAR signals and perform \gls{TDOA} calculations, enabling precise jamming signal modulation without requiring exact radar motion parameters, which are often difficult to obtain.

%\subsubsection{Monopulse Effective vs. Non-Monopulse Effective Jamming}
%\cred{I am not sure we should add this category. We will discuss methods that use monopulse radar and that is fine. Maybe we can leave the category but mention that we do not make a selection for this one.}
%
%Monopulse radars are difficult to jam, particularly in self-protection scenarios where the jammer is located on the target. These radars gather all necessary tracking information from a single pulse, making them resistant to traditional jamming techniques. In some cases, a high J/S ratio from a stand-off jammer or well-deployed decoys and chaff can be effective. However, even if range information is denied, the radar can still track by angle, making self-protection jamming especially challenging. The two main approaches to jamming monopulse radars are exploiting known radar weaknesses or manipulating the radar's angle-tracking information within a single resolution cell, with the latter being more effective.
%

\begin{figure}[t]
  \centering
  \begin{subfigure}[b]{0.24\textwidth}
    \centering
    \includegraphics[width=.85\linewidth]{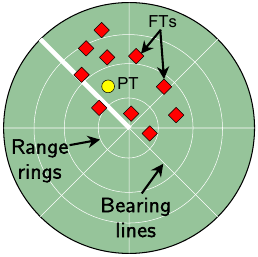}
    \caption{Search deception (e.g., multiple FTs).}
    \label{fig:search_deception}
  \end{subfigure}
  \hfill
  \begin{subfigure}[b]{0.24\textwidth}
    \centering
    \includegraphics[width=.85\linewidth]{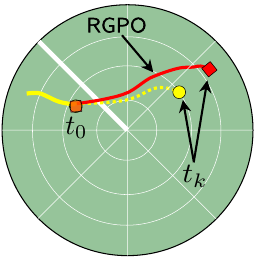}
    \caption{Tracking deception (e.g., RGPO).}
    \label{fig:tracking_deception}
  \end{subfigure}
  \caption{
    \cred{Illustration of radar deception strategies in a \gls{PPI} display (not to scale). 
    \textbf{(a)} Search deception (\Cref{sec:search}) jeopardizes lock-on by cluttering the display.
\textbf{(b)} Tracking deception (\Cref{sec:tracking}) degrades tracking accuracy or breaks lock entirely.
The spoofed and true target trajectories over the interval \([t_0, t_k]\) are shown in red and yellow, respectively.
%Multiple scans are overlaid for illustration.
  }}
  \label{fig:deception_strategies_ppi}
\end{figure}

\begin{figure}
\centering  \includegraphics[width=.92\columnwidth]{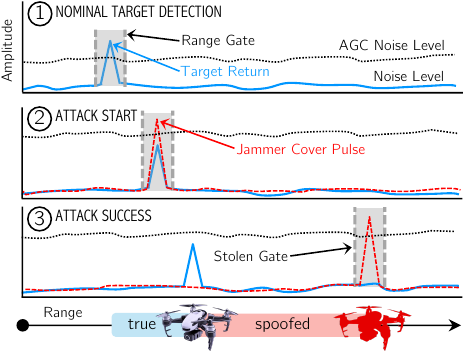}
  \caption{\cred{Illustration of a tracking gate stealing attack in the range domain, shown in three stages: \textbf{(1)} PT detection with the true return inside the range gate; \textbf{(2)} attack onset with cover pulse overlapping the target return; \textbf{(3)} successful attack where the range gate is pulled away, masking the PT return. Below are the range intervals of the moving PT and the deceptive false ranges induced by the attack.}
}
  \label{fig:amplitude_vs_delay}
  \vspace{-0.5cm}
\end{figure}

\subsection{Strategies for Radar Deception Jamming}
\label{sec:jamming_deception_taxonomy}

%\helen{Pau, I have changed the classification here to search deception attacks as first class instead of FTG-based attacks!!! let's see what u think}

% INTRODUCTION OF DRFM

%
%Deception jamming can also be implemented using simpler or older technologies, although these lack the precise control required against modern radars equipped with robust signal processing capabilities.
%
%

%\cred{Open roadmap from Google Drive to discuss whether we could add mathematical description of deception attack as an LTI system here - and other points of the paper where math could be added.}

%%A radar deception jammer can be modeled as a \gls{LTI} system applied to the radar pulse $s(t)$, resulting in the jamming signal $j(t) = (h* s)(t) = \int_{-\infty}^{\infty} h(\tau) s(t - \tau) \, d\tau$, where $*$ denotes the convolution operation. The impulse response generating $N$ pulses, each one corresponding to an \gls{FT}, can be expressed as

The signal sourced from a deception jammer, as received by the radar, can be modeled as a \cred{linear time-invariant} system applied to the transmitted radar pulse \( s(t) \). The received jamming component corresponding to the $i$-th \gls{FT} is given by $j^{(i)}(t) = \int_{-\infty}^{\infty} h^{(i)}(\tau) s(t - \tau) \, d\tau$. The impulse response of this system can be expressed as
\begin{equation}
  h^{(i)}(t) = A^{(i)} \delta (t - \tau^{(i)}) e^{j \left( 2 \pi f_D^{(i)} t + \phi^{(i)} \right)},
\end{equation}
being \( A^{(i)} \), \( \tau^{(i)} \), \( f_D^{(i)} \), and \( \phi^{(i)} \) the amplitude, delay, Doppler shift, and phase observed at the radar for the jamming component. These parameters contain false range, velocity, and angle information introduced by the deception attack.
When the jammer transmits \( N \) pulses, the signal received at the radar is given by %
\begin{equation}\label{eq:rx_signal}
   r(t) = \sum_{i=0}^{N} A^{(i)} s(t - \tau^{(i)}) e^{j \left( 2 \pi f_D^{(i)} t + \phi^{(i)} \right)} + \eta(t), 
\end{equation}
where \( i = 0 \) corresponds to the \gls{PT} return, and \( i \geq 1 \) to the \glspl{FT} introduced by the deception jammer. The term \( \eta(t) \) accounts for additive noise in reception.
%

%\cite{martone_2025_metacognitive}

%
%\cred{A deception attack may also attempt to alter the \gls{DoA} information, but this is particularly challenging when the radar system benefits from spatial diversity. Discrepancies in the \gls{DoA} across multiple sensors indicate that the signal likely originates from an \gls{FT}, as the \gls{DoA} of a \gls{PT} should be consistent with the real target location.}
%

We categorize radar deception jamming attacks according to their effects on the radar’s search and tracking operations, as depicted in Fig.~\ref{fig:taxonomy_jammming_ii}.
In search deception, we include attacks that generate echoes with \gls{FT} information typically with the aim of disrupting radar search or acquisition.
These attacks are often carried out by a jammer referred to as the \cred{FT generator}, which creates \glspl{FT} to confuse the radar system.
This is generally part of an \gls{SOJ} or \gls{EJ} strategy to enable friendly intruders to penetrate adversarial territory, but it can also be used for self-protection. 
In contrast, tracking deception aims to manipulate the perception of already established tracks rather than create new ones.
\subsubsection{Search/Acquisition Deception}\label{sec:search}
%\subsubsection{FTG-Based Attacks}
%
%%\gls{ERP}, i.e., transmitted power multiplied by antenna gain,
%\cred{These attacks } 
%
While noise jamming was traditionally used to confuse search radars, the use of radar frequency agility and diversity has made it less effective unless the jammer signal has a very high \cred{effective radiated power}. In contrast, generating multiple \glspl{FT} can be effective with a lower \cred{effective radiated power}.
%, resulting in modern radars like \gls{PD} systems often prioritizing higher sensitivity to track radar parameters. 
%
\cred{As illustrated in the \gls{PPI} display of Fig.~\ref{fig:search_deception}}, a multiplicity of echoes can clutter the radar output, thereby jeopardizing target detection. Additionally, the increased computational demand can overwhelm radar processing, resulting in significant delays~\cite{berger2003digital}.
Next, we introduce attacks that generate \glspl{FT} in the range-angle and range-Doppler domains. We also make a specific note on scene deception in \gls{SAR}, given its dedicated body of literature.
%\pau{add a sentence summarizing what is the rationale of what's coming next}
%, creating a bottleneck that delays or impairs its performance.
%
%make false targets appear credible on radar displays, such as the PPI~\cite{neri2006introduction}, cluttering the display and confusing the operator or automatic detection system, or to overwhelm and delay radar processing by creating a bottleneck.
%by generating false signals across multiple ranges and Doppler shifts, saturating the radar’s processing and delaying or preventing accurate target detection.
%to prevent target lock-on and disrupt tracking in tracking radars:
%
%
%
%\textbf{Multiple False Targets (MFT):}
%
%\paragraph{False Target Generation}
%

%

\textbf{Range-Angle Domain:} To generate an \gls{FT} in the range domain, the jammer intercepts the signal transmitted by the radar, synchronizes with its \gls{PRF}, and introduces time delays to simulate targets at varying ranges. To create realistic \gls{FT} trajectories, the jammer also synchronizes with the radar beam scanning pattern, which enables angular deception. With this, it is possible to create false echoes in the range and angular domains.
The work in~\cite{mehmood2021novel} presents a deception jamming approach against \gls{FDA} radars, creating nulls in the radar radiation pattern to conceal the true target while retransmitting time-delayed signals to generate \glspl{FT} at different ranges.
\textbf{Range-Doppler Domain:} Against \gls{PD} radars, the jammer attack exploits the reliance of these radars on coherent integration and spectral analysis for target detection.
In particular, the \glspl{FT} must remain coherent across multiple pulses. Otherwise, the energy of the jamming signal spreads across the frequency spectrum, making the attack unsuccessful. The jammer retransmits the incoming signal, adding time delay and phase modulations to simulate false Doppler shifts, thus creating false echoes in both the range and Doppler domains.
The generation of coherent \glspl{FT} within the range-Doppler space has been explored against wideband \gls{LFM} radar~\cite{iqbal2020effective,ali2022electronic}, with the study in~\cite{7858673} investigating the use of \gls{ISRJ}. 
Unlike \gls{CRDJ}, which mainly relies on replaying entire pulses with modifications, \gls{ISRJ} intermittently samples and retransmits segments of the radar pulse in a way that distorts range-Doppler processing while maintaining coherence. Compound deception jamming occurs when multiple techniques are combined, such as an attack that simultaneously employs \gls{CRDJ} and \gls{ISRJ}~\cite{li2023random}.
In addition, the work in~\cite{frazer2020deception} introduces \glspl{FT} into ground-mapping images generated by \cred{Doppler beam sharpening} radars.
Furthermore, a notable example of a velocity deception jamming attack, increasingly recognized in modern warfare as a significant threat to \gls{PD} radars, involves the introduction of micro-motion \glspl{FT}. These generate micro-Doppler shifts that simulate small, rapid movements, such as rotations or vibrations, thereby mimicking the motion of \glspl{PT} beyond simple translation~\cite{microdppler_deception}.
%In contrast, \gls{TFT} mimic the target movement along a consistent trajectory.

%
%Translational motion and micro-Doppler (m-D) modulation are combined in \cite{vehicle_target} to generate high-fidelity false target echoes.

%\paragraph{Scene Deception:}
\textbf{Scene Deception in SAR:} Although jamming \gls{SAR} systems is inherently challenging due to their reliance on long-term coherence and precise motion compensation, several studies in the literature have demonstrated successful methods for deceiving these systems.
For example, the work in~\cite{wang2019multi} employs \gls{FDA} radar to generate multi-scene attacks, showing how the number and positions of the \glspl{FT} can be controlled by tuning the \gls{FDA} antenna parameters.
The authors in~\cite{zhou2019deception} proposed a networked jamming approach for high-fidelity deception jamming against near-field \gls{SAR}, using \gls{TDOA} measurements to accurately determine radar position and generate precise modulation terms for the jamming function.
%, which allows to cover the protected object with a deceptive scene.
%
Some works focus on reducing the computational burden in the implementation of the function representing the jamming attack, which can be performed via azimuth time-domain processing~\cite{7636977} and azimuth frequency-domain processing~\cite{8309414}. Computational efficiency has also been studied for large-scene deception~\cite{yang2019large}.
Finally, the work in~\cite{tang2021evaluation} serves as a key reference, introducing a framework for evaluating the effects of deception jamming on \gls{SAR}.

\subsubsection{Tracking Deception}\label{sec:tracking}
%
%These attacks are primarily designed against tracking radars and 
%
These attacks are generally most effective after the radar has locked onto the \gls{TOI}, \cred{a scenario depicted in Fig.~\ref{fig:tracking_deception}}, and are often employed as an \gls{SPJ} strategy.
\cred{In the following, we introduce tracking gate stealing and angle tracking deception approaches.}

\paragraph{\cred{2.1)} Tracking Gate Stealing}
This category includes range and/or velocity gate-stealing strategies, such as \gls{RGPO} and \gls{VGPO}, along with their pull-in counterparts and combined range-velocity variations. These attacks target vulnerabilities in the tracking gate circuits of tracking radars.
%These include range and/or velocity gate-stealing strategies, such as \gls{RGPO} and \gls{VGPO}, as well as their pull-in counterparts (RGPI and VGPI) and combined range-velocity variations (RVGPO and RVGPI).
%
%
%
%

%
%These attacks are the main SPexploit vulnerabilities in the automatic circuits responsible for range gate tracking in tracking radars.
%
%
%, and is achieved with attack strategies like range gate pull-off (RGPO) with positive delays (distance enlargement attack) and range gate pull-in (RGPI) with negative delays (distance reduction attack).
%
%
%
%
%\vspace{-.06cm}
%\begin{enumerate}[label=\textbf{\MakeUppercase{\alph*})},leftmargin=0em,labelsep=.5em,itemindent=3.3em]

%\item Range gate pull-off (RGPO) attack: 
\textbf{Range Domain:} Radars track range by adjusting the early and late range gates to balance the energy between them. In an \gls{RGPO} attack, the jammer intercepts the radar signal and retransmits it with higher power, creating a cover pulse that manipulates the \gls{AGC} of the victim radar. This action pulls the range gate away from the true return arrival time, effectively performing a distance enlargement attack~\cite{ew_101}\cred{, as illustrated in Fig.~\ref{fig:amplitude_vs_delay}.}
    %At the same time, this lowers the automatic gain control (AGC) sensitivity and makes the true return appear weaker~\cite{fundamentals_ew}.
    %
    %Once control of the range gate is achieved, the jammer introduces delays in the retransmitted pulses. 
%
%
%
In linear RGPO attacks, the range induced by the jammer is given by~\cite{blair_benchmark_1998}
\begin{equation}\label{eq:rgpo}
    r_k^{d}=r_k^t + v_\text{po}(t_k - t_0),
\end{equation}
being $r_k^{d}$ and $r_k^t$ the \gls{FT} and \gls{PT} ranges at time step $k$, $v_\text{po}$ the attack pull-off velocity, $t_k$ the radar dwell time, and $t_0$ the attack starting time.
It is assumed that the \gls{FT} and \gls{PT} arrival times differ more than the radar range resolution, and consequently the two returns can be resolved. 
%If this was not the case, additional interference would occur and the attack would have a different form than range deception.
%
When the range gate is shifted far enough from the true target, the jammer either shuts down, forcing the radar to restart its search and lock process~\cite{blair_benchmark_1998}, or continues transmitting to maintain deception.
%
%the deception, creating an \gls{FT} remote from the real one.
    %
    %
%This tactic delays radar tracking, providing critical time for the platform carrying the jammer to evade radar-guided threats~\cite{blair_benchmark_1998}.
    %
Research efforts have focused on optimizing \gls{RGPO} strategies to improve track deception. These efforts include white-box \gls{RGPO} jamming, where the jammer has complete knowledge of the radar tracking model~\cite{9776498}, and black-box \gls{RGPO} jamming, which addresses the more realistic scenario where the jammer operates without information about the radar functioning~\cite{10032699}. In the latter case, the jammer pull-off strategy is guided by the measured deception success rate~\cite{jia_intelligent_2020}.
In the recent work in~\cite{wang2024modelling}, the effects of RGPO jamming on radar tracking are modeled, quantifying the relationship between \gls{SJR} and the maximum values of \( v_\text{po} \), providing valuable insights into optimizing RGPO strategies.
\cred{The findings in~\cite{1435790} offer insights into the vulnerabilities of \gls{DRFM}-based jamming, particularly \gls{RGPO} attacks, suggesting that anti-deception techniques can leverage the spectral artifacts introduced by phase quantization to detect and suppress jamming signals. These artifacts are central to the jamming classification approaches in~\cite{4472184, 7827908}, which we further discuss in~\Cref{sec:detection}.}
%
    %\cred{Check leading-edge tracking or anti-RGPO in 6.3.2.2 from~\cite{neri2006introduction} as anti-deception jamming technique.}
    %\item 
    %
   \Gls{RGPI} attacks operate similarly to \glspl{RGPO}, but instead of pulling the range gate outward, they pull it inward, causing the \gls{FT} to appear closer to the radar than the true target, and effectively resulting in a distance reduction attack. While \gls{RGPO} remains effective against frequency-agile radars or those with random \gls{PRF}, which are common in modern systems, \gls{RGPI} attacks are more vulnerable to these features and are thus considered less practical~\cite{6159736}.
   %However, some advanced radars remain susceptible to \glspl{RGPI}, such as pulse-compression and \gls{PD} radars.
    %, which rely on coherent pulse summation and can be deceived by partially retransmitted pulses with reduced delays, and pulse-Doppler radars, which require fixed PRF and frequency during the processing of a batch of pulses.
%
    
%
%\end{enumerate}
%
%
%\cred{Should we do a taxonomy for the jamming types and do a table with the ones we have found? If so, we will have optimization methods. Or should we just classify the type of jamming being mitigated when doing the taxonomy for antijamming methods?}
%\hl{There is some literature on RGPO attack optimization, explain from the perspective of the attacker the literature}
%
%
%
%
%
%\hl{RGPI Attacks}
%
%\textbf{VGPO/VGPI Attacks:} 
%
%https://www.youtube.com/watch?v=WAajpKTJj-8&ab_channel=RohdeSchwarz
%

\textbf{Velocity Domain:}
\Gls{VGPO} attacks are effective against \gls{PD} and \gls{CW} radars, which create velocity gates that specify the expected span of target velocities. They operate in a manner analogous to \gls{RGPO} attacks, but instead of displacing the range gate, they target the velocity gate by shifting Doppler frequency.
%
%\cred{Check section 6.3.2.3 from~\cite{neri2006introduction} for ECCM against VGPO (use of guard gates).}
%
%rely on the velocity gate to track a target’s relative speed by measuring the Doppler shift. In velocity gate pull-off (VGPO) and velocity gate pull-in (VGPI) attacks, jammers exploit this mechanism by introducing false Doppler shifts to displace the velocity gate, in a similar way to how it is done in RGPO/RGPI attacks. The jammer initially creates a false target with the same Doppler frequency as the true target, tricking the radar into locking onto the stronger false return, and gradually increases (VGPO) or decreases (VGPI) the Doppler frequency, shifting the velocity gate away from the true target. Once the velocity gate is sufficiently displaced, the jammer stops transmission, forcing the radar to restart the search.
%
Notably, the \cred{range-velocity gate
pull-off} attack can be used against \gls{PD} radars by combining the effects of \gls{RGPO} and \gls{VGPO}. %The jammer must maintain phase coherence to prevent the radar from rejecting the false signal due to inconsistencies in phase or amplitude.

\begin{figure*}
%\hspace{-1cm}                                                 
% \includegraphics[width=0.8\textwidth, height=3.8cm]{figures_TAES_FREEDOM/results/chaotic/chaotic_rmse.pdf}
\centering
  \includegraphics[width=.95\textwidth]{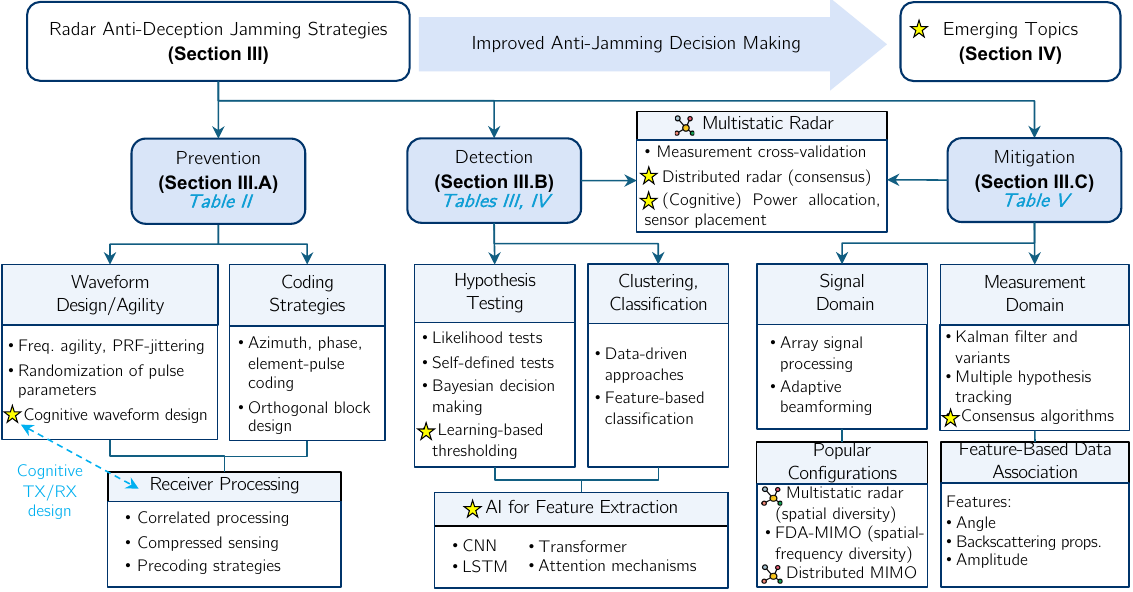}
  %\vspace{.1cm}
  \caption{\cred{Classification of radar anti-deception jamming strategies based on their functional objectives, as described in~\Cref{sec:taxonomy_anti}. The figure also highlights emerging research topics discussed in~\Cref{sec:challenge_future}. Emerging topics and multistatic radar strategies are indicated by a star and network icon, respectively.}}
   \label{fig:taxonomy_jamming_iii}
  % \vspace{-2.5ex}
\end{figure*}

\paragraph{\cred{2.2)} Angle Tracking Deception}
%Angle Tracking Attacks
%Tracking Loop Disruption Attacks
%Tracking Instability Methods
%feedback disruption
The following strategies can be used to cause a loss of lock by destabilizing angular tracking through oscillating errors, or to mislead the radar perception of the true location of the target. Some strategies involve superimposing amplitude modulation to create false angular data in sequential tracking radar systems~\cite{neri2006introduction}.
Monopulse radars are less vulnerable to jamming methods that rely on manipulating signals over multiple pulses, as they extract angular information from a single pulse~\cite{skolnik1970radar}. Therefore, we categorize angle tracking deception strategies into two types: those targeting non-monopulse radars and those targeting monopulse radars.

\textbf{Attacks Against Non-Monopulse Radar:}
Against conical scanning radars, the inverse gain technique synchronizes with the radar’s modulation and retransmits an inverted amplitude pattern, causing angular errors. 
Against \cred{lobe-on-receive-only} radars, where the jammer cannot receive or synchronize with the radar's modulation pattern, \cred{swept amplitude-modulation} uses a repetition interval modulated around the radar scan rate~\cite{sharma2020artificial}.
%
%Monopulse radars are difficult to jam because they extract all angular tracking information from a single pulse. 
%Even if range information is denied, the radar can often still track the angle, which may still be enough for weapon guidance to the target~\cite{ew_101}. 
%

\textbf{Attacks Against Monopulse Radar:}
For monopulse radars, some angular error techniques exploit the radar resolution cell, which defines the smallest angular and range separations it can distinguish. Among these, formation jamming forces the radar to track the centroid of multiple targets placed within the same resolution cell, whereas blink jamming alternates jamming signals between targets within the resolution cell to create oscillatory tracking errors. The work in~\cite{davidson2020theory} provides a theoretical analysis of blink jamming and introduces a novel variant with synchronized amplitude modulation.
Also against monopulse radar, cross-eye jamming employs two coherent sources, making it a dual-source jamming strategy, transmitting signals with matched amplitude but opposite phase. This creates a phase distortion in the wavefront, and as a result, the radar perceives the distorted wavefront as originating from a direction different from that of the TOI~\cite{ew_101}. 
In~\cite{rao2020deception}, the authors propose a cooperative dual-source jamming approach to address the challenges posed by track-to-track distributed radar fusion systems, which inherently perform better at countering deception jamming compared to other fusion methods.

\cred{\textit{\textbf{Note:} In some scenarios, search and tracking deception may exhibit common characteristics. Notably, persistent or coordinated search deception can induce tracking deviations, particularly if FTs mimic the expected behavior of the \gls{TOI}. Similarly, tracking deception may involve the injection of multiple FTs. While legacy \gls{RGPO} attacks focused on stealing the range gate to mask the PT return (as illustrated in Fig.~\ref{fig:amplitude_vs_delay}), some radars may process detections originating from both the PT and RGPO-generated signals. The latter necessitates the use of alternative protection techniques, such as those incorporating knowledge about the \gls{TOI} dynamic model~\cite{1998_Kirubarajan,calatrava2024mitigation}.}}

\section{Strategies for Radar Anti-Deception Jamming}\label{sec:taxonomy_anti}

%\helen{as shuo suggested, the text in this section should be like a textbook introduction, more generic. Try to reduce the amount of references in the text if they already appear in the tables. }

% Example of paper with very clear overlap between categories:~\cite{sun2024anti_opt}, but that is okay!

%
We classify radar anti-deception jamming strategies based on their functional objective, as illustrated in Fig.~\ref{fig:taxonomy_jamming_iii}. This includes the prevention, detection, and mitigation of deception attacks. Each category is explored in its own subsection below \cred{and may be linked to a specific stage within the radar signal processing chain, as overviewed in Fig.~\ref{fig:intro}.}
The study in~\cite{li2023random} distinguishes between passive and active anti-jamming strategies. Passive methods enhance resilience through signal processing adaptations, while active methods modify the transmitted waveform to counter deception. The latter encompasses strategies aimed at preventing deception attacks, which we discuss first.

\subsection{Prevention Strategies}\label{sec:prevention}
%\shuo{previous ``waveform diversity / design" is put here}

%\hl{Waveform Diversity/Design: Pulse Diversity, Frequency Agility, Pulse Compression, etc.). Methods Countering DRFM Functioning (anti repeater jamming)}

\begin{tcolorbox}[
    colframe=gray!50!black, 
    colback=gray!10, 
    coltitle=white,
    fonttitle=\bfseries, 
    title={\textbf{Prevention Strategies}}, 
    colbacktitle=blue!50!black, 
    boxrule=0.75mm, 
    rounded corners
]
\begin{itemize}[leftmargin=*]
%\item[+] Disrupt deception jammers by varying radar signal parameters.
    \item[+] Remain conceptually simple.
    \item[+] Force the jammer to adapt, increasing its load.
    \item[+] Some methods do not require modifications in the radar processing chain.
    \item[--] May reduce coherence in radar processing.
    \item[--] Require careful PRI/waveform management.
    %\item[--] Success depends on jammer’s sophistication.
\end{itemize}

%\begin{itemize}[leftmargin=*]
    %\item[+] Disrupts deception jammers by continuously varying radar signal parameters.
   % \item[+] Force the jammer to adapt, increasing its computational load.
   % \item[+] Some require no modifications to the radar processing chain.
   % \item[--] May reduce coherence in radar processing.
    %\item[--] Effectiveness depends on the jammer’s sophistication.
   % \item[--] Require careful PRI and waveform management.
%\end{itemize}
%
\end{tcolorbox}

\setlength{\arrayrulewidth}{0.1mm} % optional: adjust the thickness of table borders
\renewcommand{\arraystretch}{.8}  % optional: adjust row height

\begin{table*}[htp]
\small
\centering
\caption{Overview of strategies for the prevention of radar deception jamming classified according to their receiver processing approach (see~\Cref{sec:prevention}).}\label{table:jamming_prevention}
\begin{tabular}{|>{\centering\arraybackslash}m{1.5cm}|>{\centering\arraybackslash}m{0.6cm}|>{\centering\arraybackslash}m{0.6cm}|>{\centering\arraybackslash}m{7.2cm}|>{\centering\arraybackslash}m{1.3cm}|>{\centering\arraybackslash}m{4cm}|}
\hline
 \rowcolor{blue!10}
%\textbf{\begin{tabular}[c]{@{}c@{}}Receiver \\ Processing \\ Approach\end{tabular}} & \multicolumn{1}{c|}{\textbf{Ref.}} & \multicolumn{1}{c|}{\textbf{Year}} & \multicolumn{1}{c|}{\textbf{Description}} & \textbf{Deception Domain} & \multicolumn{1}{c|}{\textbf{Challenges}} \\ \hline
\textbf{Receiver  Processing  Approach} &\textbf{Ref.} &\textbf{Year} & \textbf{Description} & \textbf{Deception Domain\footnote{Domains listed are as stated in the studies; a line means none
specified.}} & \textbf{Challenges} \\ \hline
\multirow{6}{*}{\begin{tabular}[c]{@{}c@{}}Correlated \\ Processing\end{tabular}} & \cite{1435879} & 2005 & \parbox[t]{7.2cm}{Randomization of \gls{LFM} chirp parameters with a two-stage matched filter.} & Range-Doppler & \multirow{6}{*}{\parbox[t]{4cm}{$\bullet$ Coherence breaks with agile waveforms, leading to high sidelobe patterns.\\$\bullet$ Highly vulnerable, cannot handle agile deception attacks.}} \\ \cline{2-5}
 & \cite{schuerger2009performance} & 2009 & \parbox[t]{7.2cm}{Randomization of OFDM subcarrier coefficients and variation in their number.} & Range &  \\ \cline{2-5}
 & \cite{zhang_new_2013} & 2013 & \multirow{3}{*}{\parbox[t]{7.2cm}{RPIP with optimization of the initial phases aided by multi-channel processing.}} & \multirow{3}{*}{Doppler} &  \\ \cline{2-3}
 & \cite{7131032} & 2015 &  &  &  \\ \cline{2-3}
 & \cite{7485306} & 2016 &  &  &  \\ \cline{2-5}
 & \cite{s17010123} & 2017 & \parbox[t]{7.2cm}{OFDM-\gls{LFM} radar with randomly phase-modulated subcarriers.\\} & --------- &  \\ \hline
\multirow{5}{*}{\begin{tabular}[c]{@{}c@{}}Compressed \\ Sensing\end{tabular}} & \cite{7330288} & 2015 & \multirow{3}{*}{\parbox[t]{7.2cm}{RPIP with sparsity-based processing.}} & \multirow{3}{*}{Doppler} & \multirow{5}{*}{\parbox[t]{4cm}{$\bullet$ Accuracy depends on grid resolution (meshing).\\$\bullet$ High computational complexity.\\$\bullet$ Sensitive to incorrect sparsity assumption.}} \\ \cline{2-3}
 & \cite{sui2015sparse} & 2015 &  &  &  \\ \cline{2-3}
 & \cite{s18041249} & 2018 &  &  &  \\ \cline{2-5}
 & \cite{quan2018range} & 2018 & \parbox[t]{7.2cm}{Carrier frequency hopping and PRF-jitter with sparsity-based processing.} & \multirow{2}{*}{\parbox[t]{1.3cm}{\centering Range-Doppler}} &  \\ \cline{2-4}
 & \cite{li2021phase} & 2021 & \parbox[t]{7.2cm}{Block-sparse CS for range-Doppler recovery.} &  &  \\ \hline
\multirow{8}{*}{\begin{tabular}[c]{@{}c@{}}Precoding \\ Strategies\end{tabular}} & \cite{akhtar2009orthogonal} & 2009 & \parbox[t]{7.2cm}{Nearly orthogonal pulse design.} & Range & \multirow{8}{*}{\parbox[t]{4cm}{$\bullet$ Effectiveness depends on proper waveform design.\\$\bullet$ More predictable: risk of jammers exploiting the coding pattern.\\$\bullet$ Moderate computational cost.}} \\ \cline{2-5}
 & \cite{Lei2012ARE} & 2012 & \parbox[t]{7.2cm}{Full-rate orthogonal pulse block design.} & \multirow{7}{*}{\parbox[t]{1.3cm}{\centering Range-Doppler}} &  \\ \cline{2-4}
 & \cite{li2014frequency} & 2014 & \parbox[t]{7.2cm}{Frequency agility and subband synthesis.} &  &  \\ \cline{2-4}
 & \cite{abdalla2016eccm} & 2016 & \parbox[t]{7.2cm}{Orthogonal waveform design in netted radar systems leveraging signal fusion.} &  &  \\ \cline{2-4}
 & \cite{tang2020high} & 2020 & \multirow{3}{*}{\parbox[t]{7.2cm}{Group-, azimuth-, and element-pulse phase coding techniques.}} &  &  \\ \cline{2-3}
 & \cite{chen2024deceptive} & 2024 &  &  &  \\ \cline{2-3}
 & \cite{yu2023mainbeam} & 2024 &  &  &  \\ \cline{2-4}
 & \cite{li2023random} & 2023 & \parbox[t]{7.2cm}{Frequency and coding agility.} &  &  \\ \hline
\end{tabular}
\end{table*}

%
%Pulse diversity involves varying radar pulses at each PRF, with these variations being known only to the radar. 
%
We define \textit{prevention} strategies as those that dynamically adjust radar signal parameters during transmission to hinder the ability of jammers to successfully replicate the target echo. %
This is often achieved by introducing unpredictability or complexity into the radar signal design, while also implementing modifications in receiver processing to more effectively prevent deception.
In Table~\ref{table:jamming_prevention}, we categorize prevention strategies against radar deception jamming based on the receiver processing techniques employed, which are described in Section~\ref{sec:prevention:modifications_rx}.
%\vspace{-0.5cm}

\begin{figure}
\centering
  \includegraphics[width=.85\columnwidth]{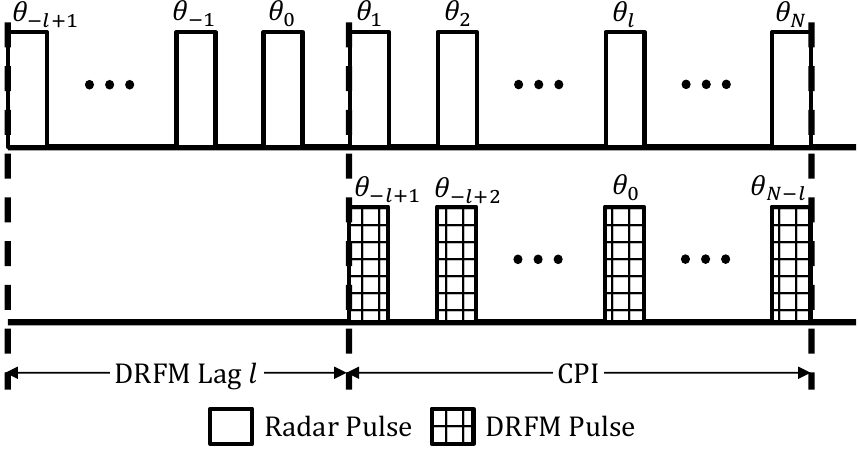}
  \caption{General scheme of pulse diversity. The radar transmits $N$ pulses within a \gls{CPI}, varying the parameter $\theta_i$ according to the pulse diversity strategy and at each pulse $i = \{-l+1, \dots, N\}$. The \gls{DRFM} jammer repeats the radar pulses with a lag of $l$ \glspl{PRI}, which results in a parameter mismatch that enables the distinction between \gls{FT} and \gls{PT}.}
  \label{fig:prevention:basics}
  \vspace{-.5cm}
\end{figure}

\begin{comment}
 %\vspace{-.8cm}
\begin{figure}[t]
\centering
  \includegraphics[width=.85\columnwidth]{figures/3_taxonomy/pulse_diversity_2.pdf}
  \caption{Illustration of an adaptive pulse diversity scheme. The signal processing block processes the received signal $r(t)$ and outputs estimates of the \gls{PT} and \gls{FT} parameters. These estimates are fed back to the design block, which updates the pulse parameters $\theta$ to prevent signal replication by the jammer.
}
  %\caption{Multi-channel processing scheme. The received echo includes a component with the signal with pulse parameters $\theta_1$ scattered by the \gls{PT}, and another one with the pulse repeated by the DRFM, with parameters $\theta_{-l+1}$. By processing the echo using the reference signal delayed by the $M$ previous \glspl{PRI}, \gls{PT} and \gls{FT} detection is achieved, as well as the estimation of \gls{FT} parameters and the DRFM lag $l$.}
  \label{fig:prevention:prevention_processing}
  \vspace{-.5cm}
\end{figure}
%
\end{comment}

\subsubsection{Pulse/Waveform Diversity}
%In Table~\ref{table:jamming_prevention}, we categorize prevention strategies against radar deception jamming based on the type of diversity employed.
%
These strategies vary the transmitted pulse (or waveform) characteristics across successive \glspl{PRI}, with the variations being known only to the radar.
Since \gls{DRFM} jammers rely on pulses from previous \glspl{PRI} to generate the jamming signal, it is challenging for them to replicate the radar pulse under these conditions. An overview of this scheme is provided in Fig.~\ref{fig:prevention:basics}.
Early examples of pulse diversity applications include the work in~\cite{blair_benchmark_1998}, which proposed using longer pulses, higher \gls{SNR} waveforms, or pulse compression to improve range resolution against \gls{RGPO} attacks. However, only works from the 2000s onward are included in Table~\ref{table:jamming_prevention} for an up-to-date review.
The prevailing body of literature focuses on waveform diversity methods, which have evolved thanks to the advancement of waveform generation technology~\cite{li2023random}. The benefits of this form of active anti-jamming are twofold: it reduces the correlation between jamming and target components, and it decreases the probability of the jammer successfully identifying and intercepting the signal~\cite{li2023random}.
Many waveform diversity strategies involve randomizing waveform parameters, such as \gls{OFDM} subcarrier coefficients or pulse initial phases, the latter using \gls{RPIP}.
In~\cite{schuerger2008deception}, the effects of employing different waveforms are evaluated, highlighting the advantages of \gls{OFDM} through frequency diversity.
%
%
%when an RGPO attack is detected or pulse compressionleverages the tracking algorithm to adaptively adjust radar parameters and reacquire the target if the signal is lost due to jamming. The tracking algorithm adapts to RGPO by continuously monitoring the relative position and velocity of the target. Since the false target created by RGPO drifts away from the true target over time, the radar can use this drift to discriminate between the real and false targets. When RGPO is detected, the radar selects waveforms that help increase the range resolution. This is typically achieved by using a longer pulse length or a waveform with higher \gls{SNR}. Additionally, pulse compression techniques can be applied to increase the radar’s range resolution, making it harder for the \gls{RGPO} to create a false target that is indistinguishable from the true target.
%
%Also recently, the work in~\cite{li2023random} 
%
The effectiveness of interpulse and intrapulse agility against deception jamming is discussed in~\cite{li2023random}. Interpulse agility (e.g., \gls{PRF} and frequency agility) prevents \gls{CRDJ} attacks from coherently replicating full pulses, whereas intrapulse agility (e.g., frequency and phase coding within a pulse) prevents \gls{ISRJ} attacks from coherently reconstructing partial pulse segments. \gls{CRDJ} and \gls{ISRJ} were introduced in Section~\ref{sec:jamming_deception_taxonomy}.

\subsubsection{Modifications in Receiver 
Processing}\label{sec:prevention:modifications_rx}
Prevention strategies typically focus on modifying the transmitted signal, but they also require corresponding adjustments in receiver processing. 
For instance, some strategies involve estimating \gls{FT} parameters, such as their range, velocity, and the lag in \glspl{PRI}, i.e., the time offset between the transmitted radar signal and the signal replayed by the jammer. Knowledge of these parameters is crucial for methods that adapt the pulse initial phases to suppress the jamming power around the Doppler frequency of the true target~\cite{zhang_new_2013,7485306}.
To estimate these parameters, multi-channel processing can be employed, with each channel used to process the signal from a different \gls{PRI}~\cite{zhang_new_2013, 7131032, 7485306}.
The information gathered about the \glspl{FT} can be leveraged to identify the jamming modality the radar is exposed to, enabling the application of additional anti-jamming measures~\cite{zhang_new_2013}.
Furthermore, entropy-based multi-channel processing can be used to relax the often unrealistic assumption that the jamming and target echoes must exist simultaneously in one \gls{CPI}~\cite{s18041249}.
As indicated in Table~\ref{table:jamming_prevention}, \cred{and illustrated by the upper feedback loop in Fig.~\ref{fig:intro}}, some pulse diversity methods enable the radar to adaptively modify the transmitted pulses based on the output of the signal processing block.
Below, we provide a discussion of the different receiver processing approaches for prevention strategies.

\paragraph{\cred{2.1)} Correlated Processing}
Conventional \gls{CP} strategies (e.g., \cred{matched filtering and coherent integration using the fast Fourier transform}) produce higher sidelobes in the correlation function when applied to signals with random or quasi-random phase, or frequency variations~\cite{s18041249}.
\gls{CP} assumes a high degree of similarity between the transmitted pulse and the local replica used for correlation. Nevertheless, when random variations in phase or frequency are introduced, they disrupt the alignment between the local replica \(s(t)\) and the received signal \(r(t)\), which is defined in Equation~\eqref{eq:rx_signal}. This misalignment causes the energy to spread across a broader frequency range, leading to higher sidelobes in the correlation function, \(R_{rs}(\tau) = \int_{-\infty}^{\infty} r(t) s(t - \tau) \, dt\). As a result, the \gls{SJR} is degraded, posing a challenge for effective target detection and jamming mitigation.
%
%In other words, although complex waveform diversity proves effective in mitigating deception jamming, its inherent multidimensional parameter variation introduces discontinuities in the phase of the received signal echo, resulting in a loss of coherence~\cite{li2023random}.

%
\paragraph{\cred{2.2)} Compressed Sensing} To address the coherence loss in \gls{CP}, some studies propose sparsity-based methods, which represent the received signal as a combination of sparse components for the \glspl{PT} and \glspl{FT}. This approach outperforms \gls{CP} in processing random signals, due to the low sidelobe characteristics of the \gls{CS} framework~\cite{s18041249}.
%
%To address the loss of coherence associated with \gls{CP}, some studies propose sparsity-based methods, where the received signal is represented as a combination of sparse components for the \glspl{PT} and the \glspl{FT}~\cite{7330288, sui2015sparse, s18041249, quan2018range}. This approach has demonstrated better performance than \gls{CP} when processing random signals, owing to the low sidelobe characteristics of the \gls{CS} framework~\cite{s18041249}.
%
%This aspect is critical to prevention strategies, as a significant number of them rely on the randomization of waveform parameters to enhance radar signal unpredictability.
%
%\Gls{CS} has proven useful in scenarios involving carrier frequency hopping and uneven sampling caused by \gls{PRF} jitter, which leads to signal phase discontinuity and, consequently, makes conventional \gls{FFT} ineffective in extracting the Doppler frequency of the target~\cite{quan2018range}.
%
%Sparsity-based methods can benefit from \gls{RPIP}, which ensures that signal components remain distinguishable when projected onto sparse dictionaries.
%
Despite its advantages, the accuracy of \gls{CS}-based parameter estimation is limited by the discretization (or meshing) of the solution space. A coarse grid resolution can lead to inaccurate parameter estimates, while a fine grid resolution significantly increases computational complexity. Additionally, \gls{CS} techniques often rely on iterative optimization algorithms, such as basis pursuit or orthogonal matching pursuit, which can be computationally demanding.
%
%Despite its advantages, the accuracy of \gls{CS}-based parameter estimation is constrained by the discretization, or meshing, of the solution space. If the chosen grid resolution is coarse, the estimated target parameters may be inaccurate, and if the grid is too fine, computational complexity increases significantly. Furthermore, \gls{CS} techniques often require iterative optimization algorithms (e.g., basis pursuit, orthogonal matching pursuit), which can be computationally expensive.
%In real-time radar systems, these methods may not be practical for fast-moving or dynamic scenarios.
%
%The authors in~\cite{li2023random} propose a.... 

%
%% Orthogonal coding
%
\paragraph{\cred{2.3)} Precoding Strategies}
These strategies enhance resilience by adding structure to the transmitted waveform, increasing signal complexity and making it harder for jammers to predict or replicate the radar pulse.
%
%For example, in group phase coding, different phase codes are applied to pulse groups~\cite{chen2024deceptive}.
%
Unlike \gls{CP}, which passively relies on coherence, precoding actively preserves it through compensation mechanisms that mitigate phase and frequency distortions introduced by waveform agility. Additionally, precoding is computationally more efficient than \gls{CS}, as it does not rely on sparse reconstruction techniques.
Despite its advantages, deterministic precoding strategies introduce a predictable structure that can be exploited by advanced repeater jammers.

\subsection{Detection Strategies}\label{sec:detection}
\begin{figure}
\centering
  \includegraphics[width=1\columnwidth]{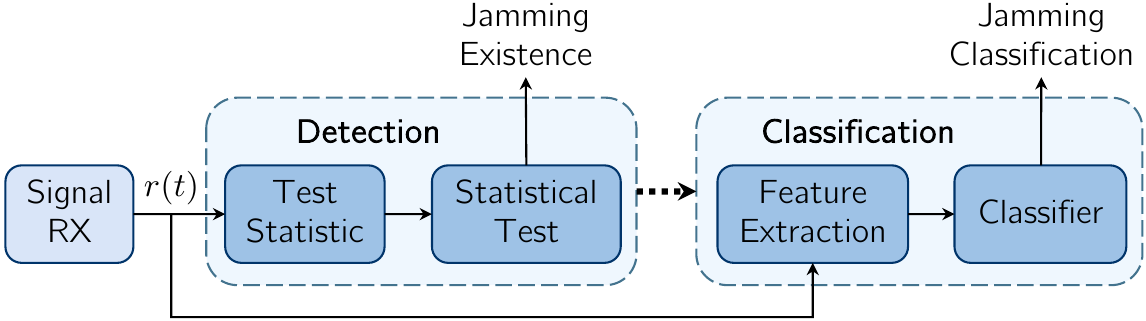}
  \caption{Illustration of the jamming signal detection and classification process. The statistical
metric is computed from the received signal for hypothesis testing, while
features are extracted for classification. Note that classification does not strictly require prior detection, \cred{as indicated by the dotted line.}}
  \label{fig:detection_diagram}
  \vspace{-.5cm}
\end{figure}

\begin{tcolorbox}[
    colframe=gray!50!black, 
    colback=gray!10, 
    coltitle=white,
    fonttitle=\bfseries, 
    title={\textbf{Detection/Discrimination Strategies}}, 
    colbacktitle=blue!50!black, 
    boxrule=0.75mm, 
    rounded corners
]
\begin{itemize}[leftmargin=*]
    \item[+] Enable identification of \glspl{FT} using statistical and threshold-based methods.
    %\item[+] Learning-based methods lessen reliance on predefined signal statistics and threshold selection.
    \item[+] Some may be integrated into existing radar architectures with minimal modifications.
    \item[--] Performance degrades under low \gls{SJR}.
    %\item[--] Many do not allow for jamming mitigation.
    \item[--] Rely on known signal statistics and threshold selection (excluding costly/overfitting learning methods).
    %\item[--] Learning-based methods require high computational resources and risk overfitting.
\end{itemize}

\end{tcolorbox}
The term \textit{detection} in this survey exclusively refers to the identification of jamming within the received signal, while \textit{discrimination} refers to distinguishing between the target and jamming echoes.
Detection typically serves as the initial step before mitigation and is often formulated as a hypothesis testing problem.
Additionally, this work explores jamming \textit{classification} (also known as jamming \textit{recognition}) approaches that not only detect jamming but also identify its type.
A block diagram illustrating the detection and classification process is shown in Fig.~\ref{fig:detection_diagram}.
Next, we review decision-making strategies for deception jamming attacks, with an overview of the relevant works provided in Table~\ref{table:jamming_discrimination}. This is followed by a focused discussion on multistatic radar systems, with a summary of the relevant works given in Table~\ref{table:detection_multistatic}, due to their prominence in the literature.
For a detailed theoretical background on detection theory, the interested reader is referred to \cite{kay2009fundamentals}.

\begin{table*}[htp]
\small
\caption{Overview of decision-making approaches for the detection/discrimination of radar deception jamming (see~\Cref{sec:detection}).}
\label{table:jamming_discrimination}
\begin{tabular}{|>{\centering\arraybackslash}m{1.78cm}|>{\centering\arraybackslash}m{0.65cm}|>{\centering\arraybackslash}m{.6cm}|>{\centering\arraybackslash}m{3.3cm}|>{\centering\arraybackslash}m{3.4cm}|>{\centering\arraybackslash}m{1.35cm}|>{\centering\arraybackslash}m{3.7cm}|}
\hline
\rowcolor{blue!10}
\textbf{Discrimination Methodology} & \textbf{Ref.} & \textbf{Year} & \textbf{Statistics / Input Data} & \textbf{Key Technique / Architecture} & \textbf{Deception Domain} & \textbf{Challenges}\\ 
\hline
\multirow{8}{*}{\begin{tabular}[c]{@{}c@{}}Hypothesis \\ Testing: GLRT\end{tabular}} & \cite{4203039} & 2007 & \multirow{8}{*}{\parbox[t]{1.85cm}{\centering Likelihood ratio}} & \multirow{4}{3.4cm}{\parbox[t]{3.4cm}{\centering Conic rejection}} & --------- & \multirow{8}{*}{\parbox[t]{3.7cm}{$\bullet$ Need for knowledge of signal statistics.\\ $\bullet$ Threshold selection.\\ $\bullet$ Sensitivity to parameter estimation.}} \\
\cline{2-3}
& \cite{4472184} & 2008 & & & & \\
\cline{2-3}
& \cite{hao2012adaptive} & 2012 & & & & \\
\cline{2-3}\cline{6-6}
& \cite{li2017adaptive} & 2017 & & & Range-Doppler & \\
\cline{2-3}\cline{5-5}\cline{6-6}
& \cite{huang2013novel}\footnote{Polarimetric processing for deception jamming discrimination was first proposed in~\cite{huang2013novel}, but the work lacked a defined decision-making approach.} & 2013 & & \multirow{3}{*}{\parbox[t]{3.4cm}{\centering Polarimetric radar system}} & --------- & \\
\cline{2-3}\cline{6-6}
& \cite{yu2019polarimetric} & 2019 & & & \multirow{2}{*}{\parbox[t]{1.35cm}{\centering Range}} & \\
\cline{2-3}
& \cite{zhang2023detection} & 2023 & & & & \\
\cline{2-3}\cline{5-5}\cline{6-6}
& \cite{7827908} & 2017 & & Phase noise examination & Doppler & \\
\hline
\multirow{4}{*}{\begin{tabular}[c]{@{}c@{}}Self-Defined \\ Test\end{tabular}} & \cite{shuangcai2011algorithm} & 2011 & Phase variance & Signal extraction and sorting & Range-Doppler & \multirow{5}{*}{\parbox[t]{3.7cm}{$\bullet$ Need for knowledge of signal statistics.\\ $\bullet$ Manual threshold selection.\\ $\bullet$ Limited generalization across radar systems and jamming types.}} \\
\cline{2-5}\cline{6-6}
& \cite{wang_2019} & 2019 & Signal power & Four-channel monopulse radar & --------- & \\
\cline{2-5}\cline{6-6}
& \cite{guo2019improved} & 2019 & Fitting residual & Bistatic radar system & \multirow{2}{*}{\parbox[t]{1.35cm}{\centering Range}} & \\
\cline{2-5}
& \cite{liu2021unsupervised} & 2021 &Homogeneity and separation score~\cite{zhao2016discrimination}, spectrum moments & Clustering analysis & & \\
\hline
\begin{tabular}[c]{@{}c@{}}Bayesian \\ Decision\end{tabular} & \cite{zhou2021feature} & 2021 & Posterior probability & Multiple-feature fusion & --------- & \parbox[t]{3.7cm}{$\bullet$ Need for effective feature selection.\\ $\bullet$ High computational complexity.\\ $\bullet$ Sensitive to low SNR.} \\
\cline{1-5}\cline{7-7}
\multirow{18}{*}{\begin{tabular}[c]{@{}c@{}}ML-Based \\Approaches\end{tabular}} & \cite{luo2021semi} & 2011 & Received time sequence & Deep belief network& & \multirow{18}{*}{\parbox[t]{3.7cm}{$\bullet$ Choice of processing domain and transformation method.\\ $\bullet$ Need for large amount of training data.\\ $\bullet$ Risk of overfitting to simulated environment.}} \\
\cline{2-5}\cline{6-6}
& \cite{lv2021radar} & 2021 & Time--frequency spectrogram & CNN + transfer learning & \multirow{2}{*}{\parbox[t]{1.35cm}{\centering Jamming Classifier}} & \\
\cline{2-5}
& \cite{kong2022active} & 2022 & Time--frequency and range-Doppler spectrogram & CNN + attention & & \\
\cline{2-5}\cline{6-6}
& \cite{bouzabia2022deep} & 2022 & CWD & CNN & --------- & \\
\cline{2-5}\cline{6-6}
& \cite{wang2023application} & 2023 & Received time sequence & Wavelet scattering network & \multirow{3}{*}{\parbox[t]{1.35cm}{\centering Jamming Classifier}} & \\
\cline{2-5}
& \cite{peng2023positive} & 2023 & Frequency response & \multirow{2}{3.4cm}{\parbox[t]{3.4cm}{\centering CNN + LSTM}} & & \\
\cline{2-3}
& \cite{wenbin2024method} & 2024 & & & & \\
\cline{2-4}
& \cite{10500322} & 2025 & Multimodal: complex envelope and kinematic information & &  & \\
\cline{2-5}
& \cite{fi15120374} & 2023 & CWD & CNN + Swin Transformer & & \\
\cline{2-5}
& \cite{10371006} & 2023 & \multirow{4}{*}{Complex signal} & Transformer &  & \\
\cline{2-3} \cline{5-6}
& \cite{rs16111989} & 2024 & & Complex-valued Transformer & Range--Doppler & \\
\cline{2-3} \cline{5-6}
& \cite{zhang2024ensemble} & 2024 & & Transformer (ensemble of subnets) & Jamming Classifier & \\
\cline{2-3} \cline{5-5}
& \cite{zhang2024weakly} & 2024 & & CNN + Transformer (weakly supervised) &  & \\
\cline{2-6}
& \cite{zhang2024intensive} & 2024 & Signal amplitude & Transformer + CFAR (dual-branch) & Range--Doppler & \\
\cline{2-6}
& \cite{lang2022jr} & 2022 & \multirow{4}{*}{STFT} & CNN + Transformer (lightweight) & Jamming Classifier & \\
\cline{2-3} \cline{5-5}
& \cite{zhu2024gcn} & 2024 & & Graph convolutional network + ViT (dual-branch) &  & \\
\cline{2-3} \cline{5-5}
& \cite{10868277} & 2024 & & Swin Transformer (distributed radar) &  & \\
\cline{2-3} \cline{5-5}
& \cite{yang2024hybrid} & 2025 & & CNN + Transformer (CvT-style) &  & \\
\hline
\end{tabular}
\vspace{1mm}
\end{table*}

\setlength{\arrayrulewidth}{0.1mm} % optional: adjust the thickness of table borders
\renewcommand{\arraystretch}{.9}  % optional: adjust row height

\begin{table*}[htp]
\small
\centering
\caption{Overview of decision-making approaches using multistatic radar for the detection/discrimination of radar deception jamming (see~\Cref{sec:detection}).}\label{table:detection_multistatic}
\begin{tabular}{|>{\centering\arraybackslash}m{1.7cm}|>{\centering\arraybackslash}m{0.6cm}|>{\centering\arraybackslash}m{0.6cm}|>{\centering\arraybackslash}m{5.1cm}|>{\centering\arraybackslash}m{1.3cm}|>{\centering\arraybackslash}m{6.2cm}|}
\hline
\rowcolor{blue!10}
\textbf{Operating Domain} & \textbf{Ref.} & \textbf{Year} & \textbf{Features/Key Techniques} & \textbf{Deception Domain} & \textbf{Challenges}\\ 
\hline
\multirow{9}{*}{\begin{tabular}[c]{@{}c@{}}Signal \\Domain\end{tabular}}
    & \cite{zhao2015signal} & 2015 & \multirow{3}{5.1cm}{Correlation test on slow-time complex envelope sequences. \\ GLRT for multiple hypothesis testing.} & --------- & \multirow{3}{*}{\parbox[t]{6.2cm}{$\bullet$ Need for knowledge of signal statistics.\\ $\bullet$ Limited to scenarios with a single jammer.\\ $\bullet$ Sensitivity to parameter estimation.}} \\
\cline{2-3}\cline{5-5}
    & \cite{zhao2016discrimination} & 2017 &  & \multirow{2}{1.3cm}{\parbox[t]{1.3cm}{\centering Range}} & \\
\cline{2-2}
    & \cite{zhao2017discrimination} & &  & & \\
\cline{2-6}
    & \cite{7377013} & 2016 & \multirow{1}{5.1cm}{Clustering analysis in amplitude ratio feature space.} & \multirow{1}{1.3cm}{\parbox[t]{1.3cm}{\centering Range-Doppler}}  & \multirow{1}{*}{\parbox[t]{6.2cm}{$\bullet$ Degraded performance in environments with dense clutter.\\  $\bullet$ Growing complexity with additional receivers.\\  $\bullet$ High computational complexity.}}\\
&&&&&\\
&&&&&\\
&&&&&\\
&&&&&\\
\cline{2-6}
    & \cite{li2019hermitian} & \multirow{1}{1cm}{2019} & \multirow{1}{5.1cm}{Binary hypothesis test using Hermitian distance of the received signal vector.} &\multirow{2}{1.3cm}{\parbox[t]{1.3cm}{\centering Range}}& \multirow{1}{6.2cm}{$\bullet$ Need for knowledge of signal statistics.\\  $\bullet$ Growing complexity with additional receivers.\\  $\bullet$ Threshold selection.}\\
&&&&&\\
&&&&&\\
&&&&&\\
\cline{2-4}\cline{6-6}
    & \cite{zhang2021target} & 2021 & \multirow{1}{5.1cm}{Feature differentiation in spatial resolution cells between PT and FT.} &  &%\\ Joint correction for the NCI detector.} &  &
    \multirow{1}{6.2cm}{$\bullet$ Need for knowledge of signal statistics.\\
    $\bullet$ Requires precise resolution cell alignment.\\
    $\bullet$ High computational complexity.}\\
&&&&&\\
&&&&&\\
\cline{2-6}
    & \cite{zhao_2021} & \multirow{1}{1cm}{2021} & \multirow{2}{5.1cm}{Correlation coefficient test for multiple-jammer scenarios. \\ Bi-quantified correlation matrix used as test statistic.} & \multirow{2}{1.3cm}{\parbox[t]{1.3cm}{\centering---------}} & \multirow{2}{6.2cm}{$\bullet$ Growing complexity with additional receivers.\\ $\bullet$ Manual threshold selection.}\\
&&&&&\\ 
\cline{2-3}
    & \cite{zhao2022cooperative} & 2022 &  & & \\
&&&&&\\
\cline{2-6}
    & \cite{liu2022anti} & 2022 & \multirow{1}{5.1cm}{Feature extraction from complex envelope sequences using CNN.} & \multirow{1}{1.3cm}{\parbox[t]{1.3cm}{\centering Range}} & \multirow{1}{6.2cm}{$\bullet$ Need for large amount of training data.\\ $\bullet$ Risk of overfitting to simulated environment.}\\
    %\multirow{2}{*}{\parbox[t]{1.3cm}{\centering Range}}
&&&&&\\
&&&&&\\
\hline
\multirow{3}{*}{\begin{tabular}[c]{@{}c@{}}Measurement \\Domain\end{tabular}}
    & \cite{bo2011study} & 2011 & \multirow{1}{5.1cm}{Feature extraction from measurement inconsistency due to velocity deception.} &Doppler & \multirow{1}{6.2cm}{$\bullet$ Unavailable jammer velocity. \\ $\bullet$ Exponential increase in hypotheses and data association burden with additional receivers.} \\
&&&&&\\
&&&&&\\
\cline{2-6}
    & \cite{8355929} & 2018 & \multirow{1}{5.1cm}{Clustering analysis based on amplitude ratio features.} & \multirow{1}{1.3cm}{\parbox[t]{1.3cm}{\centering Range}} & \multirow{1}{6.2cm}{$\bullet$ Degraded performance in environments with dense clutter.\\  $\bullet$ Growing complexity with additional receivers.\\  $\bullet$ High computational complexity.}\\
&&&&&\\
&&&&&\\
&&&&&\\
&&&&&\\
\cline{2-6}
    & \cite{huang2019joint} & 2019 & \multirow{1}{5.1cm}{Fusion-based discrimination using Doppler and spatial features.} & \multirow{1}{1.3cm}{\parbox[t]{1.3cm}{\centering Range-Doppler}} & \multirow{1}{6.2cm}{$\bullet$ Information lost in two-step detection. \\  $\bullet$ Specific design for SIMO system.}\\
&&&&&\\
\hline
\end{tabular}
\end{table*}

%\helen{m.kay: \cite{kay2009fundamentals}}
%\helen{do we want to keep the binary case or maybe we could directly introduce MHT as generic and explain for binary there are only two - ask pau about MHT abbreviation}
%\helen{M.kay detection theory: we maximize the a posterior probability  meaning maximizing $p(\bx|\mathcal{H}_i)P(\mathcal{H}_i)$}
\subsubsection{Decision-Making Approaches}
%
%\cred{In radar detection, decision-making methods are evaluated by two types of errors: Type I and Type II. Type I error occurs when a true target is incorrectly identified as a false alarm, and it is related to the probability of false alarm ($P_{fa}$). Type II error happens when a false target is missed, which is linked to the probability of detection ($P_{d}$).}
%
A binary hypothesis test can be used, where one hypothesis represents the presence of the target echo and the other represents the presence of the jamming echo~\cite{li2017adaptive, hao2012adaptive, 4203039, liu2021unsupervised, wang_2019}.
%
%The binary hypothesis test that is generally solved is
This is commonly formulated as
%
%\begin{align}\label{eq:two_hypo_test}
 %   &\text{\small$
  %  \begin{cases}
   % \mathcal{H}_0: & \text{The received signal does not contain jamming.} \\
   % \mathcal{H}_1: & \text{The received signal contains jamming.}
   % \end{cases}$} \nonumber
%\end{align}
%
%
\begin{align}\label{eq:two_hypo_test}
    &\text{\small$
    \begin{cases}
    \mathcal{H}_0: & \text{The received signal does not contain jamming.} \\
    \mathcal{H}_1: & \text{The received signal contains jamming.}
    \end{cases}$} \nonumber
\end{align}
The general solution to this binary hypothesis test involves the construction of a test statistic $\mathcal{L}(r(t))$, which is then compared to a threshold to decide in favor of one of the hypotheses as
%
%For instance, $\mathcal{L}(\cdot)$ could be constructed based following the likelihood ratio test 
%
\begin{equation}\label{eq:two_hypo_test}
    \mathcal{L}\big(r(t)\big) \underset{\mathcal{H}_0}{\overset{\mathcal{H}_1}{\gtrless}} \lambda,
\end{equation}
where $r(t)$ is the received signal defined in Equation~\eqref{eq:rx_signal}.
The decision threshold $\lambda$ can be determined theoretically or experimentally.
%
%\cred{shuo says neyman person can be used to find the threshold by fixing the false alarm rate and trying to minimize the other one or shtg like that?? check}
%
The \gls{CFAR} detector adaptively sets the threshold based on local noise statistics to maintain a fixed false alarm rate~\cite{4203039,zhang2024intensive}.
A more complex and realistic scenario arises when both the target echo and the jamming signal coexist or disappear within the received signal, requiring the application of a multiple hypothesis test~\cite{4472184, yu2019polarimetric, zhao2022cooperative, zhao2017discrimination}.
%
%\cred{This can be formulated as}
%
%\begin{align}\label{eq:two_hypo_test}
 %   &\text{\small$
  %  \begin{cases}
  %  \mathcal{H}_0: & \text{The received signal only contains noise;} \\
   % \mathcal{H}_i: & \text{The received signal contains signal from source $i$},
 %   \end{cases}$} \nonumber
%\end{align}
%
%where $i$ can be multiple and can refer to a \gls{PT} or an \gls{FT} in the scene.
%
% \begin{align}\label{eq:two_hypo_test}
%     &\text{\small$
%     \begin{cases}
%     \mathcal{H}_0: & \text{The received signal only contains target echo signal;} \\
%     \mathcal{H}_1: & \text{The received signal contains deceptive jamming signal}.
%     \end{cases}$} \nonumber \\
%     &\qquad\qquad\qquad\qquad\mathcal{L}\big(r(t)\big) \underset{\mathcal{H}_0}{\overset{\mathcal{H}_1}{\gtrless}} \lambda
% \end{align}
%
%
%\subsubsection{Decision-Making Approaches}
%\textbf{The Decision-Making Framework:} 
%
%

%Based on the uncertainty and the unknown parameters in the received signal, maximum likelihood estimation and the \gls{GLRT} are the most common methods used for decision-making. 
%
In the presence of uncertainty and unknown parameters in the statistical model, the \gls{GLRT} is a common approach for decision-making, with the \cred{maximum likelihood estimator} used for parameter estimation.
With the \gls{GLRT}, the likelihood ratio serves as the test statistic, and the decision threshold is set to control the probability of false alarm \( P_{\text{fa}} \). %This probability represents the likelihood of incorrectly detecting a target when none is present. The probability of detection \( P_d \), on the other hand, quantifies the likelihood of correctly detecting a target when one is present.
%
%\helen{what about neyman person, sprt - sequential probability ratio test??? also under some assumptions there are equivalences between glrt and np I think it is when u can use mle for parameter estimation.... should we provide a paragraph with this info??}
%
Other studies have investigated alternative self-defined tests using different metrics and thresholds, tailored to the characteristics of the jamming signal and specific application requirements~\cite{shuangcai2011algorithm, wang_2019, guo2019improved, liu2021unsupervised}.
These approaches have also demonstrated strong detection performance, either by achieving a higher probability of detection $P_\text{d}$ or by reducing $P_\text{fa}$.

Given their proven effectiveness in detection and classification tasks, \gls{ML} methods have led to a substantial body of work on deception jamming signal classification.
In many of these studies, relevant signal features are extracted and fed into data-driven classifiers, including \gls{CNN} and \gls{LSTM} architectures~\cite{lv2021radar, kong2022active, bouzabia2022deep, peng2023positive}.
%, which are subsequently trained on these features.
%
The output of the \cred{neural network}-based classifier is typically the probability of each potential deception jammer type, thereby circumventing the need for a threshold to make decisions, and with the detection task inherently built into the classification process.
In the fourth column of Table~\ref{table:jamming_discrimination}, we present the input data for the \gls{ML}-based methods rather than their statistical properties, as done for other categories such as hypothesis testing.
\cred{A more in-depth discussion of \gls{ML}-based approaches is provided in Section~\ref{sec:challenge_future}, where we examine the strengths and limitations of CNNs and explore recent efforts incorporating Transformer-based architectures as a promising direction for deception jamming recognition.}

\subsubsection{Multistatic Radar for Decision-Making}\label{sec:detection:multistatic}
\begin{figure}
\centering
  \includegraphics[width=1\columnwidth]{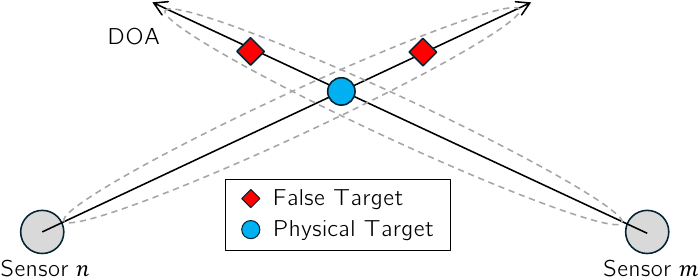}
  \caption{Illustration of the inherent resilience of multistatic radar systems against deception attacks like the \gls{RGPO} (distance enlargement) in this figure. The FT range measurements result in mismatched target positions across different sensors, while the range measurements of the PT align with a feasible target position.}\label{fig:multisensor}
  \vspace{-0.5cm}
\end{figure}

%\textbf{Multistatic Radar for Decision-Making:} 
%
%According to the jammer modulation strategy,
%
We begin by highlighting the role of multistatic radar systems in countering deception attacks, driven by the distinct behaviors of \glspl{FT} and \glspl{PT} when observed across multiple receivers.
For instance, the range bin~\cite{8355929} and Doppler frequency bin~\cite{bo2011study} of \glspl{FT} may not vary across multiple radar sensor measurements, whereas the \gls{PT} appears in different bins for each sensor.
Spatial scattering characteristics may also be leveraged, exploiting the fact that target echoes are decorrelated while deception jamming signals are highly correlated~\cite{zhao2022cooperative}.
As shown in Fig.~\ref{fig:multisensor}, even with an equidistant target, the FT range measurements across the network do not yield a feasible target position, unlike the consistent range measurements of the PT.
The spatial diversity of multistatic radar systems gives them an inherent advantage in detecting deception jamming, sparking considerable research interest. Advanced detector designs leveraging multistatic radar are summarized in Table~\ref{table:detection_multistatic}.
The works are categorized into signal domain and measurement domain approaches, where detection uses information at either the signal level or the measurement level, akin to the taxonomy presented for deception mitigation strategies in Section~\ref{sec: mitigation}.

\subsection{Mitigation Strategies}\label{sec: mitigation}
\begin{tcolorbox}[
    colframe=gray!50!black, 
    colback=gray!10, 
    coltitle=white,
    fonttitle=\bfseries, 
    title={\textbf{Mitigation Strategies}}, 
    colbacktitle=blue!50!black, 
    boxrule=0.75mm, 
    rounded corners
]
\begin{itemize}[leftmargin=*]
    %\item[+] Can suppress the impact of deception jamming at the signal and/or measurement (data) levels.
    \item[+] Wide range of techniques are applicable at the signal or measurement (data) levels.
    %\item[+] Wide range of techniques applicable at different stages of the radar processing chain.
    \item[+] Some do not rely on \gls{FT} detection.
    \item[+] Connection to Bayesian filtering enhances robustness by leveraging target history.% to mitigate the impact of erroneous measurements.
    %\item[+] Connection to Bayesian filtering improves robustness by using target history to reduce the impact of erroneous measurements on tracking performance.
    \item[--] Remain conceptually complex.
    \item[--] Require sophisticated signal processing.
   % \item[--] Rely on knowledge of noise/clutter model.
\end{itemize}
\end{tcolorbox}

%
%\shuo{``Signal-domain" and ``measurement domain" are put under this section.}
In this paper, \textit{mitigation} refers to the process of either countering the generation of deceptive measurements or mitigating their impact on system performance once they have been generated.
We refer to the former as signal domain mitigation and the latter as measurement domain mitigation, a categorization already introduced in Section~\ref{sec:detection:multistatic} in the context of multistatic radar systems for decision-making strategies.
Similarly, the work in~\cite{yang_consensus_2023} distinguishes between ``data-level" and ``signal-level" fusion mechanisms in the context of target tracking algorithms for anti-deception, which also aligns with the mitigation strategies we discuss in this subsection.
These mitigation subtypes are closely related to the two primary stages of radar detection and tracking, namely: \textit{(i)} applying the ambiguity function and matched filtering on each sensor to generate measurements, and \textit{(ii)} using these measurements to estimate target states.

 %\vspace{-.8cm}

\subsubsection{Signal Domain Mitigation}\label{sec:sigal_domain_strategy}
A substantial amount of signal domain mitigation strategies leverage multi-antenna systems, where a set of sensors observe the same signals~\cite{van2002optimum}.  
Statistical signal processing enables detection and estimation tasks on these signals, exploiting the spatial diversity of the sensor array.  
Among multi-antenna radar systems, \gls{MIMO} radar has gained significant attention. In~\cite{zhang2024game}, \gls{MIMO} is categorized into colocated and statistical \gls{MIMO}. The former exploits spatial coherence processing for waveform diversity, while the latter leverages path diversity to mitigate spatial fluctuations in target \gls{RCS}.
Beyond \gls{MIMO}, beamforming remains a fundamental technique for enhancing spatial filtering and interference suppression. This method controls the antenna beampattern to steer the main beam toward the \gls{TOI} or attenuate signals from undesired \glspl{DOA}~\cite{liu2023twenty}.
Notably, adaptive beamforming allows the array to dynamically adjust its beam pattern, optimizing signal reception while effectively suppressing noise and interference.
\Gls{DOA} estimation methods include subspace-based techniques such as \gls{MUSIC} and sparse signal recovery methods, the latter demonstrating superior performance when estimating the \gls{DOA} of highly correlated sources.
In this subsection, we examine the application of signal domain mitigation across various radar configurations, including monostatic, multistatic, and \gls{FDA}-\gls{MIMO} radar.

%
%For instance, a filtering process proposed in \cite{zhang2016range} iteratively estimates the range profiles and the range-Doppler domain locations, with results showing that both estimates converge to the vicinity of the true values. 
%
%Similarly, a coupled sequential estimation method, explored in \cite{cui2017adaptive, cui2018range}, iteratively estimates the range and Doppler frequency profile, where the \gls{PT} and \gls{FT} estimates are recursively updated.

% and a space-time four-channel radar in \cite{lei2023mainlobe} to suppress jamming.
%
Monostatic radar systems typically mitigate jamming by estimating its profile, including scatter coefficients and complex amplitudes in the range and Doppler domains. One effective approach involves recursively updating these range and Doppler profiles, progressively refining the \gls{PT} and \gls{FT} parameter estimates~\cite{zhang2016range,cui2017adaptive,cui2018range}.
\Gls{BSS} is another powerful technique for countering deception jamming, as it separates target and jamming signals into distinct channels using a separation matrix~\cite{lei2023mainlobe}. For instance, BSS has been applied in an OFDM-LFM-MIMO radar system to enhance interference suppression~\cite{gao2022mainlobe}.
While \gls{BSS} effectively isolates jamming signals, additional processing such as beamforming is still required to refine the extracted target echo and improve detection accuracy.
%
%However, the anti-jamming capability of these methods is constrained by factors such as reliance on a single signal source, low \gls{SNR}, and rapidly changing dynamics. In contrast, multistatic radar systems provide a more robust solution, as previously introduced.
%
%However, its anti-jamming capability is limited by the constraints of a single signal source, low \gls{SNR}, and the rapidly changing dynamics. In contrast, a multistatic radar system offers a more robust solution by gathering additional information about both \glspl{PT} and \glspl{FT}. This enables a globally optimized approach to counter jamming.  
%
Despite these advancements, monostatic radar systems remain inherently limited by their reliance on a single sensing node, making them more susceptible to sophisticated deception techniques, low \gls{SNR}, and rapidly changing jammer dynamics.
To overcome these limitations, radar configurations exploiting diversity have been introduced. We next discuss their role in mitigating deception attacks.

%
%Obtaining information from multiple channels and sensors is useful for improving detection and tracking performance. 
%
In modern radar systems, spatial-frequency diversity has emerged as a powerful tool to counter deception jamming by exploiting the rich information available in both the spatial and frequency domains.
This can be achieved through system configurations like multistatic radar (spatial diversity) and \gls{FDA}-\gls{MIMO} (spatial-frequency diversity). A summary of the literature on these two is provided in Table~\ref{table:spatial-frequency diversity}.
As introduced in \Cref{sec:detection}, multistatic radar systems leverage spatial diversity, where varying transmitter and receiver locations help mitigate inconsistencies in the position and velocity of \glspl{FT}. This is illustrated in Fig.~\ref{fig:multisensor}.
We have previously reviewed decision-making strategies using multistatic radar in Table~\ref{table:detection_multistatic}. Building on this, Table~\ref{table:spatial-frequency diversity} focuses on works aiming at jamming mitigation, further expanding on the role of multistatic radar in combating deception jamming.
For instance, a deception jamming suppression technique proposed in~\cite{han2022suppression} utilizes a two-radar system, consisting of one passively static radar and one actively moving radar. The method employs a statistical test based on the error covariance of angle and radial velocity measurements. Although the radar in this study is not strictly static, it is included in this category due to the spatial diversity it benefits from.
%
%\textbf{\gls{FDA}-\gls{MIMO} Radar:} 
%

\Gls{FDA} is a sensor array design that differs from traditional phased arrays by incorporating a small frequency increment across the array elements. To address the challenges associated with potential ambiguities in the range-angle dimension and time-variant beampatterns, \gls{FDA} is typically implemented in conjunction with a \gls{MIMO} system.
The additional degrees of freedom in the range domain offered by \gls{FDA}-\gls{MIMO} help separate \glspl{FT} from the \gls{PT}.
The first \gls{FDA}-\gls{MIMO} beamformer design for anti-deception jamming radars was proposed in~\cite{xu2015deceptive}, though it did not consider time delay modulation.
%
%, which has driven significant interest in using \gls{FDA}-\gls{MIMO} systems for jamming mitigation. Relevant advances and innovations in this area are summarized chronologically in Table~\ref{table:spatial-frequency diversity}.
%

It is worth noting that, due to the simplicity and widespread implementation of \gls{ISRJ} attacks, extensive research has focused on countermeasures against this type of jamming.
\Gls{ISRJ} mitigation typically involves removing it during pulse compression through two primary approaches, namely: \textit{(i)} time-frequency filtering, where a bandpass filter is designed to eliminate the \gls{ISRJ} signal from the pulse compression output~\cite{gong2014eccm, yuan2017method, chen2019band}; and \textit{(ii)} signal reconstruction, where the parameters of the \gls{ISRJ} signal are estimated and the reconstructed signal is subsequently pulse compressed~\cite{zhou2018parameter, lu2022improved, tian2023eccm}.

\setlength{\arrayrulewidth}{0.1mm} % optional: adjust the thickness of table borders
\renewcommand{\arraystretch}{1.4}  % optional: adjust row height

\begin{table*}[htp]
\small
\caption{Overview of strategies leveraging spatial and frequency diversity for the mitigation of radar deception jamming (see~\Cref{sec: mitigation}).}\label{table:spatial-frequency diversity}
\begin{tabular}{|>{\centering\arraybackslash}m{1.5cm}|>{\centering\arraybackslash}m{0.6cm}|>{\centering\arraybackslash}m{.6cm}|>{\centering\arraybackslash}m{6.6cm}|>{\centering\arraybackslash}m{1.2cm}|>{\centering\arraybackslash}m{5.4cm}|}
\hline
\rowcolor{blue!10}
\textbf{System} & \textbf{Ref.} & \textbf{Year} & \textbf{Innovation} & \textbf{Protected Domain} & \textbf{Challenges}\\ 
\hline
\multirow{4}{*}{\begin{tabular}[c]{@{}c@{}}Multistatic \\Radar \\ (Spatial\\ Diversity)\end{tabular}}
& \cite{wang2016deceptive} & 2016 & 
\multirow{1}{6.6cm}{Dual-receiver strategy that uses a coherent signal to suppress jamming at the companion receiver.} 
& Range-Doppler 
& \parbox[t]{10cm}{
    $\bullet$ Potential suppression of close-range \glspl{PT}.\\ 
    $\bullet$ Requirement for accurate estimation\\\hspace*{0em}of jamming delays.\\ 
    $\bullet$ High computational complexity.
} \\
\cline{2-6}
    & \cite{yu2020interference} & 2020 & \multirow{1}{6.8cm}{Orthogonal projection method to mitigate interference energy in the received signal.} & \multirow{30}{*}{\parbox[t]{1.2cm}{\centering Range}} & \multirow{1}{5.4cm}{$\bullet$ Alteration of signal statistics due to projection. \\ $\bullet$ Temporal misalignment of target echoes.} \\
&&&&&\\
&&&&&\\
\cline{2-4}\cline{6-6}
    & \cite{zhang2021target}\footnote{This work is listed in Table~\ref{table:detection_multistatic} as a detection technique, but here we focus on its role in mitigating jamming post-detection.} & 2021 & \multirow{1}{6.6cm}{\cred{Interference cancellation} algorithm for enhanced jamming suppression.} & & \multirow{3}{5.4cm}{$\bullet$ Need for knowledge of noise statistics.\\ $\bullet$ Precise registration of resolution cells. \\ $\bullet$ High computational complexity.} \\
&&&&&\\
&&&&&\\
\cline{2-4}\cline{6-6}
    & \cite{lan2021mainlobe} & 2021 & \multirow{1}{6.6cm}{Design of coding coefficients for \gls{MIMO} radar to discriminate \gls{PT} and \gls{FT} in the spatial-frequency domain.}&  &  
    \multirow{3}{6.6cm}{$\bullet$ Lack of clutter modeling. \\ $\bullet$ Potential suppression of close-range \glspl{PT}. \\ $\bullet$ High computational complexity.}\\
&&&&&\\
&&&&&\\
&&&&&\\
\cline{1-4}\cline{6-6}
\multirow{7}{*}{\begin{tabular}[c]{@{}c@{}}FDA-MIMO \\ (Spatial-\\Frequency \\Diversity)\end{tabular}}
    & \cite{xu2015deceptive} & 2015 & \multirow{1}{6.6cm}{First application of FDA-MIMO in radar anti-deception jamming.} &  & \multirow{4}{5.4cm}{$\bullet$ Increased false alarm rate under certain geometries.\\ $\bullet$ Data association and hypothesis explosion for large numbers of FTs. \\ $\bullet$ Need for sufficient FT samples. \\ $\bullet$ Sensitivity to mismatches in steering vectors and covariance matrices.} \\
&&&&&\\
&&&&&\\
&&&&&\\
&&&&&\\
\cline{2-4}\cline{6-6}
    & \cite{xu2018mainlobe} & 2018 & \multirow{1}{6.6cm}{Sample selection method for more accurate estimation of interference-plus-noise covariance.} & & \parbox[t]{5.4cm}{$\bullet$ Arbitrary threshold selection.\\$\bullet$ Sensitivity to mismatches in steering vectors and covariance matrices.\\}\\
\cline{2-4}\cline{6-6}
    & \cite{lan_suppression_2020} & 2020 & \multirow{1}{6.6cm}{Data-independent beamforming technique for countering mainlobe deception.} &  & \parbox[t]{5.4cm}{$\bullet$ Need for sufficient FT samples. \\$\bullet$ Lack of clutter modeling.}\\
\cline{2-4}\cline{6-6}
    & \cite{wang_main-beam_2020} & 2020 & \multirow{1}{6.6cm}{Annealing-based frequency design strategy to prevent overlap of multiple \glspl{FT}.} &  & \multirow{6}{5.4cm}{$\bullet$ Need for sufficient FT samples. \\$\bullet$ Sensitivity to mismatches in steering vectors and covariance matrices.}\\
&&&&&\\
&&&&&\\
\cline{2-4}
    & \cite{liu2022robust} & 2022 & \multirow{1}{6.6cm}{Robustness under multipath via MUSIC and Capon-based covariance reconstruction.} & & \\
\cline{2-4}
    & \cite{zhang2024mainlobe} & 2024 & \multirow{1}{6.6cm}{Adapted GoDec algorithm for robust mainlobe jammer suppression in multipath environments.} & & \\
\cline{2-4}\cline{6-6}
    & \cite{luo2024sparse} & 2020 & \multirow{1}{6.6cm}{Sparse Bayesian learning for accurate parameter estimation in covariance reconstruction.} & & 
    \multirow{1}{5.4cm}{$\bullet$ Need for sufficient FT samples. \\$\bullet$ Potential mismatch in parameter priors.}\\

\hline
\end{tabular}
\end{table*}

\subsubsection{Measurement Domain Mitigation}\label{sec:measurement_domain_mitigation}
There is also a body of literature on measurement domain mitigation using multi-radar systems. For example, the work in~\cite{bo2011study} proposes a \gls{SIMO} architecture with one transmitter and three receivers placed at varying distances from the target. By analyzing the Doppler frequencies of the three receivers, the system can determine whether the target is a \gls{PT} or an \gls{FT} generated by a velocity deception attack.
Additionally, the authors in~\cite{sun2024anti_opt} propose a power optimization strategy for a multi-radar system performing \gls{MTT} to combat deception jamming, integrating the deception range into the augmented target state. Notably, the posterior \gls{CRB} is used to build the objective function under resource constraints. Similarly, the study in~\cite{zhang2022joint} also leverages the posterior CRB to guide optimization in a distributed MIMO radar system to increase tracking accuracy. Consistent with much of the literature on measurement domain mitigation, both works primarily focus on target tracking, which we discuss next.

%
%However, this approach is specifically designed for \gls{SIMO} radar. %systems and may suffer from efficiency losses due to the two-step discrimination process.
%proposed an deception jamming suppression algorithm based on consecutively \gls{FT} identification in range and Doppler domain.
%\shuo{This is the only reference I found for ``multistatic", `` measurement domain", ``not only detection".}
%
%
%
%
%\textbf{Target Tracking Strategies:}\label{sec:ss_strategy}
%\shuo{This part should be merged into ``measurement domain strategies"}
%radar handbook skolnik: \cite{skolnik1970radar}
%

%
%Effective target tracking also plays a crucial role in mitigating deception jamming.
%
%In the following, we discuss target tracking algorithms specifically designed for deception jamming mitigation.
%
Radar systems can use mechanical or electronic beam steering systems to track targets, as seen in \gls{CW} radars or phased arrays, which adjust the beam to follow target movement. Alternatively, tracking can occur after detection~\cite{1999_li}, and typically involves estimating the target’s position and velocity based on measurements such as range, Doppler shift, and angle, obtained during successive scans.
This process is challenging, particularly since the motion state of the target can change unpredictably during maneuvers. As such, tracking strategies must balance model knowledge with real-time measurements to maintain accuracy. Methods like the \cred{Kalman filter} or \gls{MHT} are commonly applied in this context.

%Multiple hypothesis tracking (MHT) is the leading method for addressing the data association (DA) problem in modern target tracking systems~\cite{1102177,1263228} since it considers multiple hypotheses about the TOI state and updates them as new measurements are received.
%
%In radar systems, tracking occurs on two distinct levels: signal-level tracking and data-level tracking. At the signal level, the radar directly processes the received signals and tracks either the real target or a false target, such as one created by range deception jamming (e.g., RGPO). Here, the radar system can be tricked into locking onto the false target as it manipulates the range gate, causing the radar to lose track of the real object. On the other hand, 
%Data-level tracking involves post-detection algorithms like Multiple Hypothesis Tracking (MHT), which operate on the detections provided by the radar. 

%
\Gls{MHT} considers multiple hypotheses about the target state and updates them as new measurements are received~\cite{1102177,1263228}.
This multi-level approach enhances decision-making by refining potentially ambiguous radar outputs through sophisticated data association strategies. These strategies effectively manage uncertainty and determine which measurements correspond to which source, such as \glspl{PT}, \glspl{FT}, and clutter or false alarms.
\Gls{MHT} can discard unlikely tracks by assuming that the target dynamics and the type of deception attack present in the scene are known.
When \gls{FT} measurements follow patterns that do not match the expected dynamics, false tracks are filtered out.
Moreover, if signal features are available, even when the jamming track mimics the behavior of the \gls{PT} and successfully follows its dynamics, it may still be possible to mitigate the jamming component.
%

%
%\cred{Even in the cases where the method relies on signal feature extraction, it is included here given the main focus on measurement processing.}
%

Implementing these methods in real-time requires significant computational resources to ensure the filter can track both \glspl{PT} and \glspl{FT} in dynamic environments and that the number of hypotheses is controlled. However, a growing body of literature is exploring efficient implementations of \gls{MTT} algorithms, and applying these techniques to anti-deception jamming represents an important direction for ongoing research.
%
%\helen{Can I use newline/enter if I am within a subsection with bold header?}
%
Although data-level tracking algorithms provide track continuity, deception jamming identification still requires an extra level of decision-making process. This can be guided by heuristic methods, such as assuming that larger ranges correspond to \glspl{FT} in the presence of \gls{RGPO} attacks. For instance, in~\cite{1998_Kirubarajan} they reduce the association probabilities of measurements at farther ranges.
However, these assumptions can lead to significant errors or track loss, especially in the context of \gls{RGPI} attacks or false alarm measurements~\cite{766953}. 
%
%\cred{(It is not clear whether the cited paper~\cite{766953} says that this happens or whether in this paper we observe track loss in the presence of RGPI attacks.)}. 
%
Some \gls{MHT}-based approaches use signal features such as amplitude information to improve deception identification, with the amplitude difference between cover and target return pulses proving particularly informative for enhancing tracking accuracy~\cite{hou_multiple_nodate, 7377013}.
The fact that \gls{FT} measurements often have nearly identical angles to \gls{PT} measurements may also be leveraged~\cite{lan_suppression_2020}, allowing for the identification of deception based on small angular differences between measurement pairs~\cite{1995_slocumb, 1999_li}.
%
%Another method detects deception by using shared angles of false and real target measurements to improve tracking via measurement fusion, although true target identification is not performed~\cite{1999_li}.
%
%
The study in~\cite{4472184} takes advantage of a spatial feature where the steering vector of the deception jammer aligns on a cone centered around the \gls{TOI} steering vector.
%
%Furthermore, the amplitude difference between cover and target return pulses has proven informative, potentially enhancing tracking accuracy~\cite{7377013}.
%
Additionally, in~\cite{rao2011joint}, a target discrimination method combines continuous tracking with recognition to differentiate \glspl{PT} and \glspl{FT} by analyzing their backscattering properties. However, the method's high computational complexity, particularly due to the use of the nine-dimensional \cred{extended Kalman filter}, presents challenges for practical implementation.

Nevertheless, the methods outlined above are based on low-order statistics and may be insufficient when jamming signals and true targets share similar features.
To address this, leveraging information from multiple radars can help reduce state uncertainty~\cite{6178069, li2019hermitian}. As an example, in~\cite{yang_consensus_2023}, a consensus algorithm is employed to enhance tracking accuracy in distributed radar networks under deception jamming.
The study in~\cite{rs16142616} enhances tracking accuracy through a collaborative resource management strategy in a distributed \gls{MIMO} system, and uses the predicted conditional \gls{CRB} as a performance metric for joint delay and Doppler estimation to guide radar resource scheduling.
Finally, the work in~\cite{calatrava2024mitigation} relies solely on motion state information by embedding knowledge of the spatial behavior of \gls{RGPO} attacks into the clutter model assumed by the tracker through the use of \cred{random finite set} theory for \gls{MTT}~\cite{mahler_statistical_2007}. \cred{Random finite sets} allow modeling of measurement sets with variable cardinality~\cite{9353973} and are particularly useful in the presence of detection uncertainty and false alarms~\cite{vo_bayesian_2008}.
\begin{figure}
\centering
  \includegraphics[width=1\columnwidth]{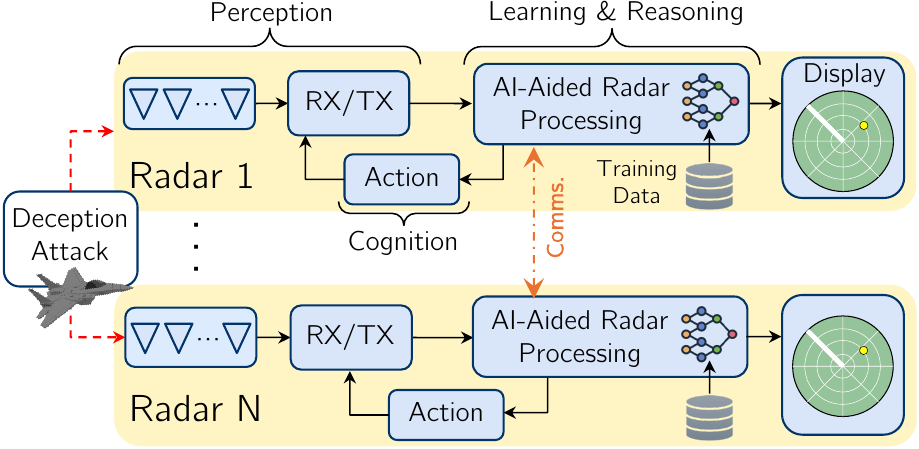}
  \caption{\cred{Overview of emerging radar architectures integrating distributed processing through multi-radar cooperation, cognitive decision-making, and AI-enabled signal processing to enhance protection against deception jamming attacks.}}\label{fig:emerging}
  \vspace{-0.5cm}
\end{figure}

\section{Emerging Topics}\label{sec:challenge_future}
As radar deception techniques continue to evolve, the need for effective countermeasures has become more critical than ever. This section examines recent advancements in distributed, cognitive, and AI-enabled radar systems that offer new avenues for anti-deception jamming. \cred{As illustrated in Fig.~\ref{fig:emerging}, these architectures leverage interconnected components for perception, learning, and reasoning, aiming to improve situational awareness.} Several methods from Section~\ref{sec:taxonomy_anti} are revisited here in light of their relevance to these emerging technologies. For each subtopic, we also outline promising research directions to guide future innovation.

\subsection{Distributed Radar}

Distributed systems allow each radar node to update its target belief by incorporating data from neighboring nodes rather than relying on a central coordinator~\cite{chalise2022distributed}. This approach enables more flexible and adaptive responses, even in the presence of deception jamming, as demonstrated in recent studies.
For example, the optimization of power allocation policies to improve tracking performance under power constraints in distributed radar networks is explored in~\cite{sun2024anti}. Additionally, \gls{FT} identification methods that leverage information from both active and passive nodes within the network are proposed in~\cite{yang_consensus_2023}.
For a detailed discussion on algorithms and implementation challenges in distributed radar networks, readers may refer to~\cite{chalise2022distributed}, though deception jamming is not the focus. Moreover, the trade-offs between centralized and distributed radar architectures are outlined in Table~\ref{table:centralized_distributed}.

\begin{table*}[ht]
\centering
\small
\caption{Comparison of centralized and distributed radar systems.}
\label{table:centralized_distributed}
\begin{tabular}{@{}ccc@{}}
\toprule
\rowcolor{blue!10} 
\textbf{Aspect} & \textbf{Centralized Radar} & \textbf{Distributed Radar} \\ 
\midrule
\textbf{Performance}  & Typically higher due to global data access. & May be suboptimal due to partial data fusion. \\ 
\textbf{Scalability}  & Limited by the central node. & Highly scalable and adaptable. \\ 
\textbf{Robustness}   & Vulnerable to single-point failures. & More resilient to node failures. \\ 
\textbf{Complexity}   & Centralized processing at the fusion center. & Processing distributed across local nodes. \\ 
\textbf{Privacy}      & More susceptible to centralized data leaks. & Better privacy through local data retention. \\ 
\textbf{Latency}      & Higher due to data transmission delays. & Lower for localized processing. \\ 
\bottomrule
\end{tabular}
\end{table*}

%
%\helen{This paragraph talks about cooperative deception --> link to cooperative antideception as a response!!}
%
%Cooperative jamming involves a coordinated effort among multiple entities that synchronize their actions and share information for stronger deception. 
%
%Especially in the context of distributed radar systems, cooperative deception jamming has gained increasing relevance in recent years.
%
%Especially in the context of distributed radar, Cooperative deception jamming presents a significant challenge for anti-deception jamming strategies. 
%

%
Recent advances in cooperative deception jamming techniques have demonstrated the ability to disrupt radar networks, regardless of their spatial diversity. In~\cite{yang_consensus_2023}, cooperative deception is described as a method that can infer the network topology and generate \glspl{FT} for multiple radars simultaneously, overcoming the misalignment issues highlighted in Fig.~\ref{fig:multisensor}. In such cases, the spatial diversity provided by multiple radars is insufficient to distinguish between \glspl{FT} and \glspl{PT}, highlighting the need for an additional layer of protection.
To address this,~\cite{yang_consensus_2023} proposes a countermeasure against deception jamming by integrating passive radars and employing a consensus-based data-sharing mechanism, enhancing target tracking resilience in the presence of cooperative jamming.
Furthermore, the relevance of distributed \gls{MIMO} radar networks is discussed in~\cite{zhang2024game}. The authors analyze the interaction between a MIMO network and the jamming source within the context of game theory, which we will explore in the next subsection. In this framework, each MIMO radar aims to maximize the mutual information between the jammer and its echo while minimizing the transmitted power.
Distributed \gls{MIMO} is also used for cooperative resource scheduling to improve \gls{MTT} performance in~\cite{rs16142616,zhang2022joint}, and for the joint estimation of \gls{PT} and \gls{FT} parameters through factor graph representation and an iterative message passing algorithm in~\cite{yu2021message}.

Research on enhancing the resilience of distributed radar systems against deception jamming remains an active area of study. A key improvement for existing methods is the incorporation of more realistic assumptions, such as addressing challenges like non-overlapping fields of view among sensors, as highlighted in~\cite{yang_consensus_2023}.
Moreover, the work in~\cite{zhao2022cooperative} tackles the challenge of registration errors in distributed radar systems, which arise from unsuccessful synchronization between sensors, leading to misaligned measurements.
Furthermore, while distributed systems keep data local, they are not entirely secure, as the information shared between nodes remains vulnerable to exposure.
Considering this, we highlight privacy-preserving anti-deception jamming as an interesting future research direction.
Federated learning, used in other fields like \gls{GNSS} jamming classification~\cite{wu2025federated}, could potentially be applied to radar systems to enable decentralized, privacy-aware mitigation strategies.

% Most current investigations focus on centralized radar networks, where local measurements are broadcast to a fusion center for simultaneous processing and estimation. 
% %
% This distributed approach provides greater flexibility, reduced communication burden~\cite{HE202021}, and enhanced robustness, motivating the development of a new data-level fusion anti-deception jamming method for distributed radar networks. 
% %
% Nevertheless, some works have explored enhanced robustness in centralized fusion architectures~\cite{6582732}.
%

\subsection{Cognitive Radar}\label{sec:cognitive_anti_deception}

%The anti-deception jamming technique have raised wide interest for keeping radar systems' recognition and tracking performance under interference. However, the deception jamming techniques have also been developed for decades to protect the target's information under radar system coverage. 
%
%As adaptive jamming strategies, such as those explored in \cite{10106133}, become increasingly sophisticated, understanding their evolving nature is crucial for developing effective \gls{ECCM}.
%
%The techniques of adaptive jamming style decision, known as \gls{RJDM}, are investigated in \cite{10106133}. The \gls{RJDM} allows the jammer to follow the optimum jamming strategy and generate the best countermeasurements based on the state of the protected target and potential threaten. This working pattern is also recognized as \textit{cognition}. To counter the intelligent jammers, the \textit{cognitive} anti-jamming strategy is needed to adaptively change its anti-jamming strategy or change the parameter of the radar systems. For example, a power allocation optimization method is proposed in \cite{sun2024anti_opt} to improve the anti-jamming and tracking performance. A cognitive waveform and filter design technique is developed in \cite{ge2021joint} against deception jamming.
%\cite{zhang2023cognitive}
Cognitive radar decision-making~\cite{10106133} allows jammers to dynamically adjust their tactics based on environmental factors through the so-called perception-action cycle~\cite{gurbuz2019overview}.
%
%The feedback mechanism by which cognitive radar systems adjust their parameters and strategies based on environmental factors can be referred to as the perception-action cycle~\cite{martone2021closing}. 
%
As intelligent deception attacks evolve, the emphasis on cognitive anti-deception jamming strategies is increasing in response.
%
%For instance, recent work has proposed power allocation optimization methods~\cite{sun2024anti_opt} to enhance anti-jamming and tracking performance, 
%
The work in~\cite{7485306} is an early example of cognitive waveform design to counter velocity deception jamming by adapting the initial pulse phases to form frequency stopbands around jammed true targets, thereby increasing the \gls{SJR}.
In~\cite{wen2019cognitive}, the perception phase involves transmitting a high \gls{PRF} waveform to detect the jammers and estimate their \gls{DOA}, while the action phase involves adjusting the radar transmit pattern to create notches around the jammer directions.
A more recent cognitive prevention strategy is presented in\cite{ge2021joint}, where the authors propose a joint design of the transmit waveform and receive filter under \gls{SNR} constraints.
%
%
%
%Moreover, other anti-deception techniques, such as multistatic radar systems and FDA-MIMO designs, mentioned in Section~\ref{sec:taxonomy_anti}, have been targeted in jamming development.
%
%For example, a phase modulation method tailored to combat deception jamming in multistatic radar systems is proposed in~\cite{9966941}, while a novel \gls{FTG} technique for creating effective deception on FDA radar is presented in~\cite{tan2020novel}.
%
%%Additionally, several of the anti-jamming techniques discussed in Section~\ref{sec:taxonomy_anti} incorporate cognition. Within the prevention category, specifically under pulse diversity, a cognitive waveform design strategy has been proposed 
%
%
%
Additionally, the concept of metacognitive radar, recently presented in~\cite{martone_2025_metacognitive}, leverages human learning principles to enhance system adaptability. This framework balances exploration (learning new strategies) with exploitation (optimizing existing strategies), enabling the radar system to adapt more effectively to changing environments. Notably, its potential use to counter deception jamming is still to be explored and represents a promising area for future research.

Evolving cognitive strategies closely align with the principles of game theory, which offers a structured framework for understanding the dynamic interplay between \gls{ECM} and \gls{ECCM}\cite{he2024game, ibrahim2024game, wang2024design}. Examples of game-theoretic anti-jamming strategies include the optimization of polarization in transmission\cite{zhang2019game}, and the joint beamforming and power allocation in multistatic radar~\cite{he2021game}. Another study proposes the introduction of an additional transmitter-receiver pair transmitting false information to divert the jammer's power away from the real communication channel~\cite{nan2020mitigation}. This scenario is modeled as a leader-follower game, where the system (leader) allocates power first, and the jammer (follower) adjusts its jamming strategy based on the signals from both the real and fake channels. For a comprehensive review of the role of game theory in defense systems, see~\cite{ho2022game}, though it should be noted that deception jamming is not the main focus.
%
%Game theory provides a framework for modeling strategic interactions between rational decision-makers, each aiming to optimize their outcome while countering the strategies of their opponent.
%
%Remarkably, the competitive interaction between \gls{ECM} and \gls{ECCM} aligns with game theory, where deception and anti-deception measures evolve in response to each other~\cite{he2024game, wang2024design}.
%
As noted in~\cite{geng2023radar}, even when accounting for a smart (strategic) jammer, the jamming models assumed by most studies remain overly simplistic, employing static games or dynamic games with single-round interactions. In contrast, real-world \gls{EW} scenarios involve multiple rounds of interaction with imperfect information. Consequently, a clear direction for future research is to develop more realistic jammer models within the game-theoretic framework to better capture the nuances of deception attacks.
%
%This gap is particularly significant in the field of anti-deception jamming, where the literature is limited and further investigation is essential.
%

%In the context of radar and jamming systems, the radar system and the jammer can be viewed as players in a non-cooperative game. Further research can delve into this adversarial relationship at an abstract level, modeling the decision-making processes between radar and target within the framework of anti-jamming strategies.

% Nash Equilibrium?

% This adversarial relationship encourages both of jamming and anti-jamming to be more advanced. This competitive interaction between each other brings the concepts of \textit{game theory}, in which the anti-deception jamming and deception jamming are evolving against each other\cite{he2024game, wang2024design}. More research can be conducted in this direction in an abstract level to reflect the adversarial relationship between the radar and target in (anti)jamming decision making.  

%\helen{add this: \cite{ nan2020mitigation, zhang2024game}. very generic game theory in defence applications reviewL\cite{ho2022game} }

\subsection{AI-Enabled Radar}

%Building on \gls{ML} in cognitive radar design, \gls{RL} offers a powerful approach for developing anti-deception jamming strategies.
%
%In this context, the \gls{RL} agent adaptively selects actions that increase resilience in the presence of jamming.
%
Building on \gls{ML} in cognitive radar design, \gls{RL} offers promising and innovative research paths. In the context of anti-deception jamming, the \gls{RL} agent can adaptively select actions to enhance system resilience in the presence of deception threats.
An early example of this paradigm is presented in~\cite{kang2018reinforcement}, where frequency-agile strategies introduced in Section~\ref{sec:prevention} are extended by adapting frequency hopping patterns to maximize rewards and improve the \gls{SJR}.
These strategies primarily focus on adjusting key parameters, such as power allocation, sensor placement, and detection thresholds, based on feedback from rewards~\cite{jiang2023intelligent}. %This enhances the ability of the radar to discriminate between \glspl{PT} and \glspl{FT}, thereby improving anti-jamming decision-making~\cite{jiang2023intelligent}.
Additionally, it is worth noting that \gls{RL} has also been used to enhance deception jamming from the perspective of the attacker. For instance, the work in~\cite{zhang2023cognitive} draws inspiration from the pheromone mechanism of ant colonies to enhance exploration and convergence speed in jamming decision-making.

\cred{
As shown in Table~\ref{table:jamming_discrimination}, modern deception jamming detection increasingly relies on \gls{ML}, with \gls{CNN}-based approaches being particularly prominent~\cite{lv2021radar,kong2022active,bouzabia2022deep,peng2023positive,wenbin2024method,10500322}. CNNs can automatically learn discriminative features from data without relying on expert domain knowledge~\cite{fi15120374}. Depending on how the input is formatted, CNN-based methods may be classified as one-dimensional, i.e., using raw echo sequences, or two-dimensional, i.e., relying on spectrograms, most commonly derived from the \gls{STFT}.
Furthermore, CNNs can inform adaptive decision-making, as demonstrated in~\cite{wang2025unified}, which integrates CNN-based jamming recognition with a policy network to select optimal anti-jamming waveforms based on passive radar inputs.
While they have proven effective, CNNs exhibit key limitations~\cite{zhang2021end}, including: \textit{(i)} degraded performance with limited training data, which has been addressed in~\cite{9079547,9320987}, with the latter leveraging generative adversarial networks~\cite{creswell2018generative} and variational autoencoders; \textit{(ii)} difficulty in handling hard samples; and \textit{(iii)} limited exploitation of the signal’s multidimensional structure in the time–frequency domain~\cite{10577225}.
Notably, the small data challenge has also been tackled through other learning strategies, such as transfer learning~\cite{lv2021radar,wang2023few} and domain adaptation techniques~\cite{xiao2024pspnet}.
Hard samples refer to radar echoes that are particularly difficult to classify, often arising from high jamming-to-signal ratios, significant target–jammer overlap, or complex jammer modulations~\cite{li2021novel,huang2016active}. The nonlinear effects introduced by the latter are especially difficult to model explicitly and often require adversarial sample generation to simulate challenging training scenarios~\cite{wang2017fast}, thereby improving detector robustness. To address these issues, the approach in~\cite{10577225} incorporates time–frequency domain information for enhanced feature extraction, an attention mechanism to focus on informative regions, and a generative adversarial training framework.
}

%
%
%

%Over the past three years, Transformer-based models have made significant strides in the recognition of deception jamming signals. These architectures are particularly attractive for their ability to model global dependencies, effectively capturing relationships between distant elements within an input sequence. This property is especially advantageous in radar signal analysis, where deceptive patterns often manifest non-locally across time or frequency domains. Unlike recurrent models such as

\cred{
Radar signals inherently form time sequences, making them particularly well-suited for sequential models such as LSTMs and the increasingly popular Transformer-based architectures~\cite{10371006}, as reviewed in Table~\ref{table:jamming_discrimination}. A representative instance of sequential modeling is provided in~\cite{10500322}, where a multimodal fusion framework integrates a \gls{CNN} to extract complex envelope features from the received signal, and an attention-enhanced bidirectional \gls{LSTM} to capture temporal dependencies from kinematic time-series data. The latter derives position information of the \gls{TOI} using echo delay, azimuth, and elevation angle measurements.
Over the past three years, Transformer-based models have achieved notable advances in the recognition of deception jamming signals~\cite{fi15120374,10371006,rs16111989,zhang2024ensemble,zhang2024weakly,zhang2024intensive,lang2022jr,zhu2024gcn,10868277,yang2024hybrid}.
Their strength lies in modeling global dependencies, which is particularly valuable in radar signal analysis, where deceptive patterns may emerge non-locally across time or frequency. Unlike recurrent models such as LSTMs, Transformers process entire sequences in parallel, avoiding fixed-step recurrence and offering increased robustness to non-stationary attack patterns, including variations in \glspl{PRI}. This results in improved generalization across diverse jamming strategies. For instance, the approach in~\cite{zhu2024gcn} demonstrates superior performance compared to random forest and both one-dimensional and two-dimensional CNN baselines across a range of jamming conditions. The architecture employs a dual-branch design: one branch uses a graph convolutional network to extract spatial features, while the other leverages a Transformer to capture global dependencies. The outputs from both branches are then fused through a feature integration module for final classification.
}

\cred{
A detailed overview of Transformer-based approaches for deception jamming recognition is presented in Table~\ref{table:jamming_discrimination}.
In general, the Transformer-based pipeline for deception jamming recognition consists of the following stages: \textit{(1)} transformation of the raw radar signal using time–frequency analysis methods such as the \gls{STFT}~\cite{yang2024hybrid,10868277,zhu2024gcn,lang2022jr } or the \gls{CWD}~\cite{fi15120374}; \textit{(2)} signal pre-processing, including operations like normalization and denoising~\cite{fi15120374}; \textit{(3)} input encoding, where the processed signal is converted into a suitable format for the Transformer through tokenization or feature embedding; \textit{(4)} processing by a Transformer architecture such as Vision Transformer (ViT)~\cite{zhu2024gcn}, Swin Transformer~\cite{fi15120374}, or Convolutional Vision Transformer (CvT)~\cite{yang2024hybrid}; and \textit{(5)} a final classification stage that outputs the predicted jamming label.
}
\cred{
Overall, the integration of \gls{AI} into radar anti-deception is a promising yet still maturing research area. The rise of Transformer-based architectures~\cite{wrabel2021a} has opened new possibilities, while underscoring the continued need for innovation in both deception and countermeasure strategies. A key challenge for Transformer-based methods is the lack of publicly available datasets: to the best of our knowledge, no real radar data or standardized synthetic datasets have been released. Existing studies using Transformer architectures rely on custom simulations that are not shared, limiting reproducibility and fair comparison across different studies. Additionally, the difficulty of labeling radar jamming signals in real-world settings has been emphasized~\cite{zhang2024weakly}.}

\section{Conclusion}\label{sec:conclusion}
This paper provides a comprehensive and up-to-date review of strategies designed to protect radar systems from deception jamming. We begin by laying the foundation with key ECM/ECCM concepts, followed by an in-depth analysis of radar deception jamming strategies, categorized into search and tracking deception. Search deception primarily involves \cred{the generation of \glspl{FT} to overload} or confuse the radar’s search and acquisition processes, while tracking deception includes gate-stealing attacks like RGPO/RGPI, and angle deception.
A major contribution of this work is the development of a comprehensive taxonomy for anti-deception jamming strategies, structured into three functional categories: prevention, detection, and mitigation. Prevention strategies aim to hinder the jammer’s ability to introduce false information; detection strategies alert the system to deception and may classify the type of attack; and mitigation strategies focus on reducing or suppressing the impact of jamming.
Within the mitigation category, we distinguish between signal-domain mitigation, which suppresses deceptive measurements at the signal level, and measurement-domain mitigation, which addresses the impact of deceptive measurements on state estimation.
Finally, we highlight key challenges and promising future research directions, particularly the integration of distributed and cognitive radar systems, alongside AI-driven techniques. These include game-theoretic approaches, Transformer-based models, and reinforcement learning, all of which hold significant potential for advancing radar anti-jamming capabilities.

%
%\appendices
%\section*{Appendix: List of Acronyms}
%\glsaddal
\setglossarystyle{list} 
\glssetcategoryattribute{acronym}{format}{\glsentrylong}
\printglossary[title={List of Abbreviations} ]

%\section*{APPENDIX}

%Appendixes, if needed, appear before the acknowledgment.

% \section*{ACKNOWLEDGMENTS}

% Research was sponsored by (...)

%\bibsection*{REFERENCES}

\bibliographystyle{ieeebib}% use plainnat to see year and name of authors, use ieeebib for TAES style, to see [] instead of () comment \usepackaga{natbib} at the top of this script
\bibliography{ofdmradar_adversarial, survey, dj_mht}
\vspace{-0.5cm}

 % HELENA
 \begin{IEEEbiography}[{\includegraphics[width=1in,clip,keepaspectratio]
 {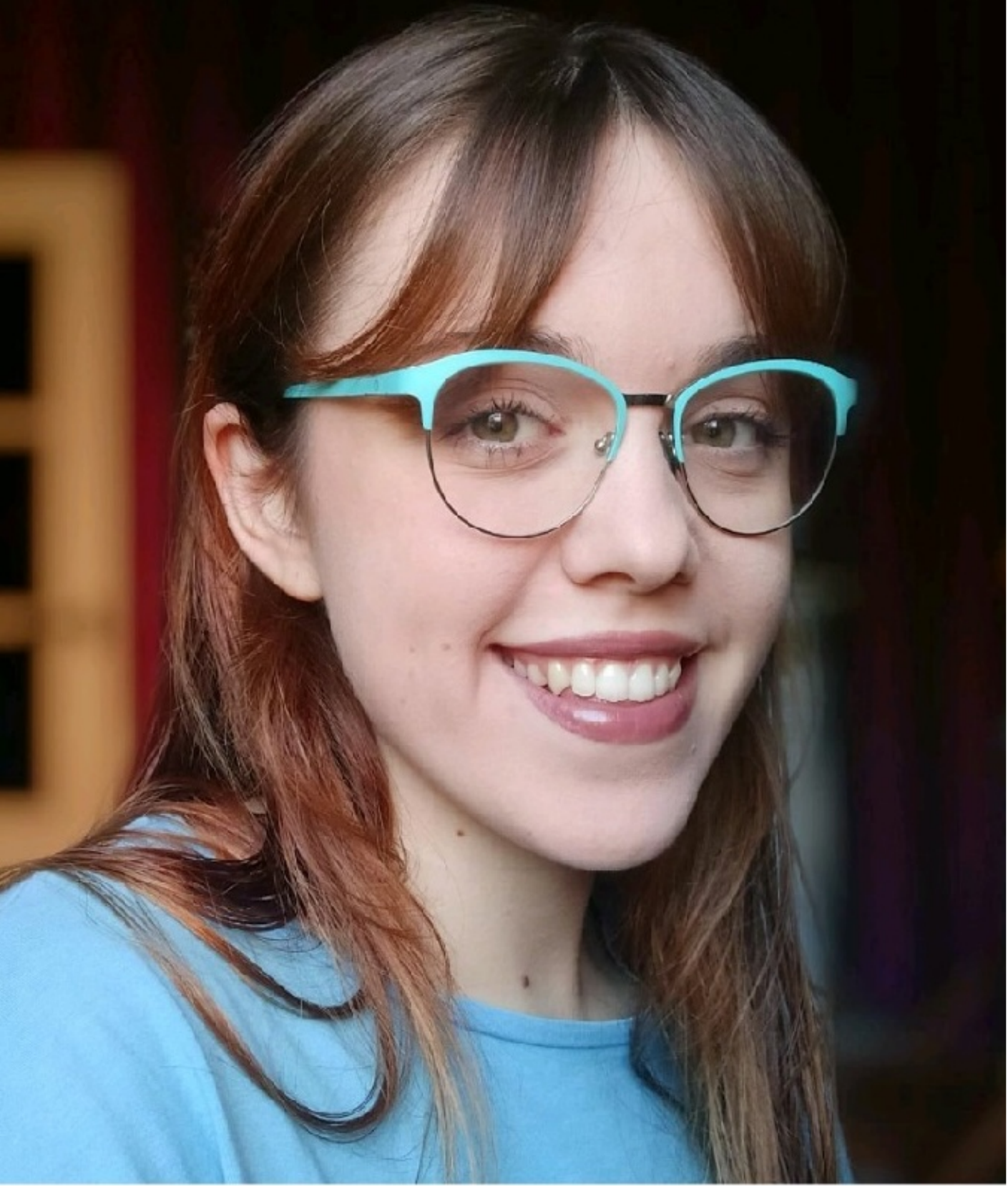}}]{Helena Calatrava}{\space}received her BS and MS degrees in Electrical Engineering from Universitat Politècnica de Catalunya (UPC), Barcelona, Spain, in 2020 and 2022, respectively. She is currently a PhD candidate in Electrical Engineering at Northeastern University's Information Processing Laboratory in Boston, MA. Her research focuses on statistical signal processing, multiple target tracking, multi-agent systems, and robust signal processing to improve resilience in GNSS and radar applications. \end{IEEEbiography}%
 % SHUO
\begin{IEEEbiography}[{\includegraphics[width=1in,height=1.25in,clip,keepaspectratio]{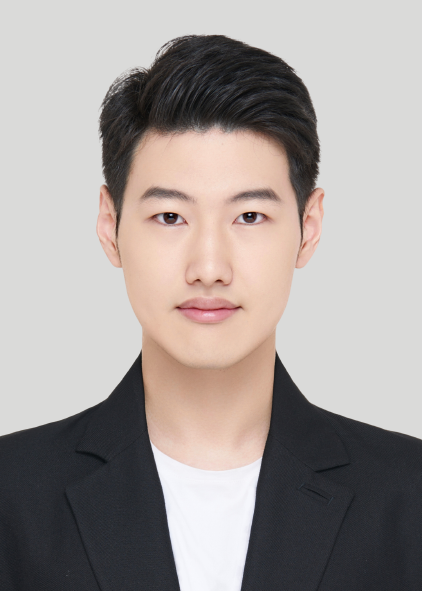}}]{Shuo Tang}{\space} received the BS degree in mechanical engineering from China Agricultural University, China and the MS degree in mechanical engineering from Northeastern University, Boston, MA, in 2018 and 2020, respectively.
He is currently a PhD candidate in electrical and computer engineering at Northeastern University. His research interests include GNSS signal processing, positioning and navigation, sensor fusion and physics-based learning.
\end{IEEEbiography}%

 % PAU
 \begin{IEEEbiography}[{\includegraphics[width=1in,height=1.25in,clip,keepaspectratio]{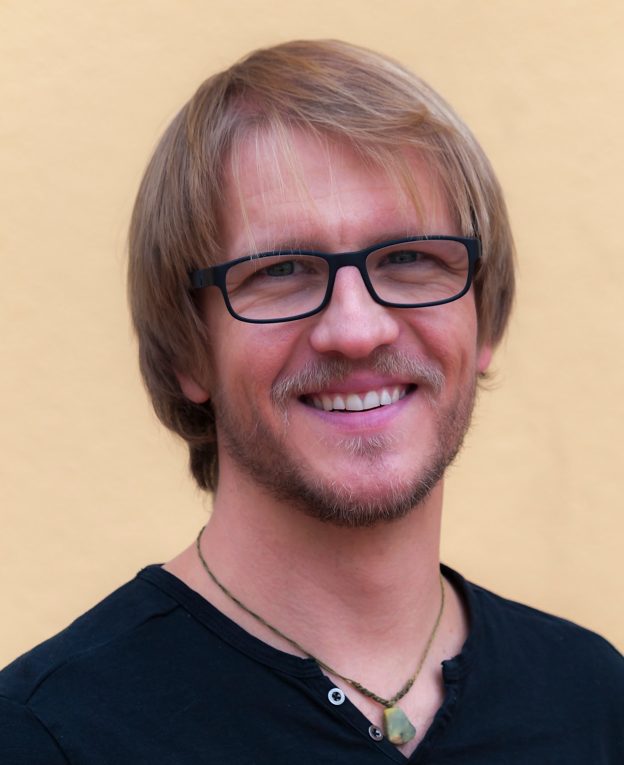}}]{Pau Closas}(Senior Member, IEEE),
 is an Associate Professor in Electrical and Computer Engineering at Northeastern University, Boston MA.
 He received the MS and PhD in Electrical Engineering from UPC in 2003 and 2009, respectively, and a MS in Advanced Mathematics from UPC in 2014. He is the recipient of the EURASIP Best PhD Thesis Award 2014, the $9^{th}$ Duran Farell Award for Technology Research, the $2016$ ION's Early Achievements Award, $2019$ NSF CAREER Award, and the IEEE AESS Harry Rowe Mimno Award in $2022$. 
 His primary areas of interest include statistical signal processing and machine learning, with applications to positioning and localization systems. He is EiC for the IEEE AESS Magazine, volunteered in multiple editorial roles (e.g. NAVIGATION, Proc. IEEE, IEEE Trans. Veh. Tech., and IEEE Sig. Process. Mag.), and was actively involved in organizing committees of a number of conference such as EUSIPCO (2011, 2019, 2021, 2022), IEEE SSP'16, IEEE/ION PLANS (2020, 2023), or IEEE ICASSP'20.
 \end{IEEEbiography}

\end{document}